\documentclass[preprint]{aastex}

\newcommand{\mdot}{M{\raise 1.5ex\hbox{\hskip-6pt$\mathchar"201$}\kern0.2em}_{acc} } 

\shorttitle{Angular Momentum Evolution}
\shortauthors{Wolff et al.}

\begin{document}


\title{
THE ANGULAR MOMENTUM EVOLUTION OF 0.1-10 M$_{\odot}$ STARS 
FROM THE BIRTHLINE TO THE MAIN SEQUENCE
}

\author {S. C. Wolff, S. E. Strom}
\affil{NOAO; 950 N. Cherry Ave.; Tucson AZ 85719}
\email{swolff@noao.edu}
\author {L. A. Hillenbrand}
\affil{Caltech; MS 105-24; Pasadena CA 91125}


\begin{abstract}
Projected rotational velocities (\textit{vsini}) have been measured 
for a sample of 145 stars with masses between 0.4 and $>$10 M$_{\odot}$ 
(median mass 2.1 M$_{\odot}$)
located in the Orion star-forming complex. These measurements 
have been supplemented with data from the literature for Orion 
stars with masses as low as 0.1 M$_{\odot}$. The primary finding
from analysis of these data is that the upper envelope of
the observed values of angular momentum per
unit mass (J/M) varies as M$^{0.25}$ 
for stars on \textit{convective} 
tracks having masses in the range \ensuremath{\sim}0.1 to \ensuremath{\sim} 3 
M$_\odot$. This power law extends smoothly into the 
domain of more massive stars (3 to 10
M$_\odot$), which in Orion are already on the 
ZAMS.  This result stands in sharp contrast to the properties of
main sequence stars, which show a break in the power law and a sharp
decline in J/M with decreasing mass for stars with M 
\texttt{<} 2 M$_{\odot}$.
A second result of our study is that this break is seen already
among the PMS stars in  our Orion sample that are on \textit{radiative}
tracks, even though these stars are only a few million years old. 
A comparison of rotation rates seen for stars
on either side of the convective-radiative boundary shows that stars
do not rotate as solid bodies during the transition from convective
to radiative tracks.

As a preliminary demonstration of how observations
can be used to constrain the processes that control early stellar
angular momentum, we show that the broad trends in the data can
be accounted for by simple models that posit that stars:  1) 
lose angular momentum before they are deposited on the 
birthline, plausibly through star-disk interactions; 
2) undergo
additional braking as they evolve down their convective
tracks; and 3) are subject to core-envelope decoupling during
the convective-radiative transition.
\end{abstract}




\section{INTRODUCTION}

Substantial progress has been made over the past several years 
in characterizing the evolution of angular momentum of low mass 
(M \texttt{<}  0.5 M$_{\odot}$) pre-main-sequence (PMS) stars as they evolve 
toward the main sequence (e. g. Herbst, Bailer-Jones, \& Mundt 
2001; Rebull 2001 and references therein). In this paper, we 
report new observations of angular momentum in intermediate-mass 
PMS stars in the mass range 0.4-3 M$_{\odot}$ and ZAMS stars for a coeval
population at masses $>$4 M$_{\odot}$, ranging up to $>$10 M$_{\odot}$.
Our first goal is to establish for stars 
in this mass range the values of angular momentum (J) for 
stars with ages of \texttt{<} 1 Myr, which we will take as an estimate 
of the initial angular momentum.  Our second goal is to compare
these values with older stars to examine
how angular momentum changes as a function of time.

To place our observational results in context, we will adopt what has become
the standard framework for early stellar evolution, i.e. that stars  
acquire a significant fraction of their final mass through rapid 
accretion via disks. If
protostars end the phase of rapid accretion quickly, that is 
on a time scale that is short relative to the time scale for 
contraction, then it is possible to derive a relationship between 
mass and radius and to use this relationship to define a locus 
of points in the temperature-luminosity plane where the rapid 
accretion stops, stars become visible, and begin their quasi-static 
contraction to the main sequence. This locus is the birthline 
(Stahler 1983).  Stars may then continue to accrete material at a low rate 
as they evolve from the birthline down convective tracks. 

Stars that form as a result of accretion of high angular momentum 
material through a disk should in principle be rotating at nearly 
the breakup speed (Durisen et al. 1989), but the typical rotational 
velocities of the youngest visible stars are instead observed 
to fall an order of magnitude or so below this critical velocity 
(Stauffer \& Hartmann 1986; Rhode et al. 2001). 
The low rotational velocities have 
been explained by positing that stars are locked to their surrounding 
disks via magnetic fields and that the disk applies a braking 
torque (Uchida \& Shibata 1984; K\"{o}nigl 1991; Shu et al. 1994). 
Therefore, a second key element of the overall framework is that 
accreting protostars lose angular momentum through interaction
with an accretion disk before they arrive 
at the birthline.

In order to account for subsequent changes in angular momentum 
and surface rotation as PMS stars evolve from the birthline
onto the main sequence, 
a number of additional factors must be taken into account: 1) 
continued loss of angular momentum via coupling to disks at the 
lower accretion rates characteristic of T Tauri stars; 2) changes 
in the moment of inertia as a result of changes in stellar radius 
and internal structure; and 3) possible decoupling within the 
star between rotation of its core and rotation of its envelope.

In the sections that follow, we first describe observations designed 
to determine the angular momenta of extremely young intermediate 
mass stars. These data provide an estimate of the initial angular 
momentum as a function of mass. We then compare these observations with 
older stars and with 
the above theoretical framework in order to learn whether these basic 
ideas can account for the observed changes in
stellar rotation as a function of time and mass. 

\section{OBSERVATIONS}

\subsection{Sample Selection}

An ideal sample for our purposes is one that (1) contains a statistically 
significant number of young stars having M \texttt{>} 0.5 M$_{\odot}$; and 
(2) includes objects born in multiple star-forming episodes so 
that stars of identical mass can be observed in a variety of 
evolutionary states. The Orion OB association is one of the nearest 
star-forming regions that satisfy these criteria. The 
ages of the stars in this region range from less than 1Myr to 
10 Myr (Walker 1969; Warren \& Hesser 1978; Brown, deGeus, \& deZeeuw 1994;
McNamara, et. al. 1989).

A sample of stars brighter than magnitude B=14 with proper motion 
membership probabilities larger than 50\% was selected 
from the studies of McNamara et al. (1989), van Altena et al. 
(1988), and Jones \& Walker (1988). This sample is spread over 
a diameter of about 2.75 degrees located within the Orion Ic/Id 
association.

For each sample object, we require photometric and spectroscopic 
data sufficient to determine: (1) the star's location in the 
HR diagram; (2) its projected rotational velocity; (3) whether 
or not the star is surrounded by a circumstellar accretion disk, 
which might ``lock'' a star to a fixed rotation period (e.g. K\"{o}nigl 1991); 
and (4) whether or not a star is a short-period spectroscopic 
binary in which each member's rotational velocity could be tidally locked 
to the orbital velocity. 

We obtained spectra of resolution and wavelength coverage sufficient 
to determine spectral types and also to measure projected rotational 
and radial velocity values.
We also obtained near-infrared photometry (JHK), which in combination 
with derived spectral types and extant optical photometry enables 
estimates of excess emission above photospheric levels and serves 
as a diagnostic of circumstellar accretion disks. We culled from 
the literature optical (UBVI) photometry as well as spectral 
types and \textit{vsini} for a few stars for which we obtained IR 
photometry but no spectroscopy. 

The databases for our sample are given in Tables 1 and 2. In 
general, Table 1 includes the observations, while Table 2 includes 
such derived quantities as visual extinction (A$_V$), effective temperature
(T$_{eff}$), luminosity (L$_{bol}$), radius (R), moment of inertia (I),
mass (M), age (A), and infrared excess ($\Delta(H-K)$). The 
entries in the tables are described in the sections that follow.

\subsection{Spectroscopic Measurements}

High-resolution (R\ensuremath{\sim}20,000) spectroscopic data were obtained 
in November 1992 and January 1993 with the Hydra multi-object 
spectrograph mounted on the KPNO 4-m telescope. The ``heart of Hydra" 
is a robotic fiber-positioning device that permits 
the observer to locate as many as 97 fibers at locations within 
a 45 arc minute field to sub-arcsecond precision. Typically, 
we were able to optimize fiber placements to enable simultaneous 
observation of 40 to 60 object spectra and 30 to 40 sky spectra. 
Priority in the assignment of fibers was given 
to Orion stars that had been shown to have excess emission at 
near infrared wavelengths by our JHK photometry. 
Two echelle settings with a spectral range of \ensuremath{\sim 150 \AA} 
were used. The first was centered at \ensuremath{\lambda}4450 and included 
He I 4471 and Mg II 4481. The second was centered at \ensuremath{\lambda}4861 
and included the H\ensuremath{\beta} line. Spectra centered 
on the He/Mg region enable rotational velocity measurements for 
stars spanning a wide range in spectral type, while those centered 
on H\ensuremath{\beta} provide a complementary measure of rotational 
velocity for stars of type A5 and later. 

Our observing strategy was to obtain the He/Mg and the H\ensuremath{\beta} 
settings on different nights of the observing runs in order to 
enable the identification of spectroscopic binaries via radial 
velocity variations. On each night, the data were obtained in 
two steps, first by taking a series of exposures with the telescope 
and fibers aligned on our target positions, and then by offsetting 
the telescope by \ensuremath{\sim}5 arcsec to obtain a ``sky'' exposure. 
This procedure (as opposed to the more standard practice of achieving 
sky subtractions by averaging several ``sky'' fibers) was necessitated 
by the large and spatially variable background of the Orion Nebula, 
which was assessed empirically via comparison between sky fibers 
from ``on-target'' to ``offset'' telescope positions.

We also obtained spectra of a grid of relatively bright stars 
to be used as rotational velocity standards. For spectral types 
later than A0 these were stars in the Pleiades cluster (Anderson, 
Stoeckly, \& Kraft 1966) and field stars (Soderblom, Pendleton, 
\& Pallavicini 1989) for which accurate rotational velocities 
are known. For hotter stars, the standards were taken from Slettebak, 
et al. (1975).

The Hydra data were reduced with the IRAF script {\bf dohydra} 
developed by Frank Valdes at NOAO. The first step in the reduction 
procedure requires removal of a low spatial frequency pattern, 
which is superposed on the spectra and results from the summed 
contributions of light emerging from the object and sky fibers 
and then scattered by the optical train. The scattered light 
background is removed using the IRAF task {\bf apscatter} 
by (1) measuring its contribution outside the regions occupied 
by object and sky spectra;
(2) fitting a two-dimensional surface to the scattered light 
pattern; and (3) subtracting the surface fit. 

The next steps involve extraction of the object and sky spectra 
and wavelength calibration. The latter step is accomplished by extracting
spectra of a Th-Ar arc source taken with the same fiber configuration 
used to record stellar and sky spectra and deriving individual 
dispersion solutions for the arc spectrum corresponding to each 
fiber. The resulting wavelength vs. pixel fits are then applied 
to the intensity/pixel relationship for the corresponding object 
or sky spectrum. The final step requires subtraction of a sky 
spectrum from each object spectrum. Typical sky corrections 
were less than 10\% of the stellar signal even for the faintest 
stars in our sample.
The resulting typical signal-to-noise per resolution 
element is between 30 (for the faintest stars) and 150 (for the 
brightest stars). 

Spectral types were derived via comparison with our grid of spectral 
standards. For stars of types mid-G and earlier, our classifications 
enable placement on a standard sequence to within \ensuremath{\pm}1 subclass. 
For stars of later type (which are also fainter on average), 
the uncertainties are somewhat higher, perhaps \ensuremath{\pm}2 subclasses 
on average, owing to the difficulties inherent in classifying 
spectra for which 
the number of decisive classification features contained within 
the relatively short wavelength span sampled is small  and for which
the S/N ratio is modest (30/1).

Rotational and radial velocities were measured with a cross-correlation 
technique. By cross-correlating an object spectrum of unknown 
line width with that of a template spectrum of known line width 
(usually taken to be a slowly rotating star with \textit{vsini} smaller 
than the instrumental resolution), one can measure the width 
of the resulting cross-correlation function. The width of this 
function is proportional to broadening
introduced by rotation and thus to \textit{vsini}. Tonry \& Davis (1979) 
originally developed this technique in an astronomical context 
in order to derive velocity dispersions in galaxies. The IRAF 
{\bf rv} package contains a script {\bf fxcor} 
which incorporates the Tonry and Davis formalism. 

Our observations span a very wide temperature range, and the 
appearance and particularly the density of absorption lines changes 
dramatically over this spectral range. Therefore, we have used 
spectral type dependent techniques to estimate apparent rotational 
velocities. 

For stars with spectral types earlier than A5, where the number 
of lines contained within our spectral windows is too small to 
enable accurate \textit{vsini} and \textit{v(radial)} 
measures from cross-correlation techniques, projected rotational velocities 
were obtained via visual comparison with artificially broadened 
slowly rotating templates of comparable spectral types (from O9-A5) 
and crosschecking with standard stars 
having known rotational velocity. Radial velocities were 
obtained from measurements of line centroids and then placed 
on the internal HYDRA system for the later-type stars from comparison 
of radial velocity values derived from line centroids and cross-correlation 
peak measurements for a selection of stars in the spectral type 
range A5 to F5. Our estimated uncertainty in \textit{vsini} 
for stars A5 and earlier is 50 km/s, 
for typical \textit{vsini} values 100-200 km/s.  Our estimated uncertainty 
in \textit{v(radial)} is 10 km/s. To be listed as 
a candidate spectroscopic binary, a star with spectral type A5 
or earlier must have a velocity which differs by 20 km/s from 
the cluster mean.

For stars later than A5, the first step in our procedure was 
to fit a function to the stellar continuum level and subtract 
this function from both the object and the template spectra. 
We then filtered the object and template spectra with a function 
tailored to reject low spectral frequencies (i.e. contributions 
to the power spectrum deriving from broad blends of atomic or 
molecular features). For the F, G, and K stars in our sample 
for which the rotational velocities were typically \textit{vsini} 
\texttt{<} 100 km/s, we rejected frequencies corresponding to features 
with widths 200 km/s or greater; the expectation that the line 
widths were less than 200 km/s was verified by visual inspection 
of the spectra. For the A stars in our sample, we used a filter 
that rejected only those features with widths 400 km/s or greater. 
The object and template spectra were then cross-correlated. By 
fitting a gaussian to the resulting function, we determined both 
its peak (which measures the relative radial velocity of object 
and template stars) and its full-width half-maximum (which provides 
a measure of the rotational velocity of the object). This procedure 
closely follows that described by Hartmann et al. 1986, who used 
a similar technique to derive rotational velocities for solar-type 
PMS stars.

In practice, we chose a template star, 
SAO 217014 (G5 V; \textit{vsini} \texttt{<} 5 km/s), which was used as the 
cross-correlation template for all object stars with spectral types F0-K5. 
For stars with 2 or more independently measured spectra, 
the agreement between the full-width half-maxima (FWHM) of the 
best-fit gaussians to the cross-correlation function is generally 
better than 5\% for FWHM \texttt{<} 60 km/s and better than 10\% for 
broader lines.  

The relationship between FWHM and projected rotational velocity 
for the stars later than A5 is established via comparison with 
published \textit{vsini} values for the Pleiades stars included in 
our Hydra sample. Figure 1 shows the relationship between FWHM 
and {\it vsini} for the 34 stars used to establish the functional 
relationship between these quantities. With the exception of 
three obviously discrepant points, the scatter about the adopted 
calibration curve is \texttt{<}10\% at all values of \textit{vsini}.
We estimate  
typical precision of 20 km/s and 3 km/s for rotational and radial
velocities, respectively, for stars of spectral type A5 and later. 

Our measurements of \textit{vsini} are listed in Table 1 along with 
previous measurements by other authors.  A number of the stars 
have been measured in other studies, but in order to make the 
data set as homogenous as possible, we have adopted (see Table 
2) our own measurements of \textit{vsini} rather than averaging our 
measurements with those in the literature. We have, however, 
supplemented our data by using values of \textit{vsini} from the literature 
for a few stars for which we had photometry (see below) but no 
spectroscopy. In adopting values from the literature we have 
confined ourselves to a few large datasets (see the notes in 
Table 2) that had some stars in common with the current survey 
and for which there was good agreement with our measurements.

Our data set also permits a search for spectroscopic binaries. 
Because Hydra records spectra for a large sample of cluster members 
simultaneously, we can compare the velocities derived from the 
location of the cross-correlation peak for each star with the 
mean velocity for all stars in a given Hydra exposure. We tested 
this procedure for our Pleiades cluster standard stars of spectral 
types A5 and later and find that the scatter about the mean is 
typically \ensuremath{\pm}2.5 km/s (1 \ensuremath{\sigma}) for these high S/N 
spectra. For our Orion data, the scatter is \ensuremath{\pm}4 km/s. Stars 
in our sample with spectral types A5 and later that deviate 
by more than 2\ensuremath{\sigma} (8 km/s) from the mean in any of our 
Orion sample Hydra exposures are tagged in column 15 of Table 
1 as \textit{candidate} spectroscopic binaries, along with the observed 
range in velocity in km/s. We emphasize that confirmation of 
{\it candidates} as true spectroscopic binaries (as 
opposed to field star interlopers sharing the proper motion but 
not the radial velocity of the Orion association) will require 
additional observations.

The 12 candidate binaries include convective as well as radiative
stars and both high and low mass objects.  The binaries do not occupy a
special location in any of the diagrams relating angular momentum
to mass.  Accordingly, we do not treat them separately in the
discussion that follows.
\subsection{Infrared Observations}

In planning our overall program, we believed that IR measurements might lead to
discovery of a significant number of disks around the stars in our sample.  The
quantity $\Delta$(H-K), the difference between reddening-corrected H-K color
and underlying photospheric color, provides a
strong discriminant between disked and diskless stars (Hillenbrand, et al.
1992; Edwards et al. 1993; Hillenbrand et al. 2004, in preparation).
  
Near-infrared (JHK) observations were obtained for several hundred 
stars in the Orion Ic/Id association that met the proper motion 
criteria defined above. The observations were made in November 
and December, 1991 with the OTTO photometer mounted on the KPNO 
50-inch telescope. The data were taken using a standard beam-switching 
pattern (source/sky/sky/source) through a 15 arcsec beam with 
a 30 arcsec throw. Calibration was performed with the standards 
of Elias et al. (1982), thus placing the data on the CIT photometric 
system within color terms of order 5\% (Kenyon 1988). Some 
measurements were adopted from the SQIID and NICMASS imaging photometry 
described by Hillenbrand, et al. (1998). The infrared data are 
summarized in Table 1.

Values of $\Delta$(H-K) are listed in Table 2; stars with 
$\Delta$(H-K) $>$ 0.1 mag
are considered disk candidates.  Our results show that only 20\% of the
stars in our sample show excesses this large and only 14 of these stars lie
on convective tracks where we might expect the stellar magnetic fields
required for disk-locking to be present.  Given the small number of stars
with excesses, it was impossible to effect a statistically meaningful
comparison of N(vsini) between stars which show, and those which lack, 
inner accretion disk signatures, 
given the broad intrinsic range of rotational velocities
characterizing stars in this mass range.  While we have made no further use
of the IR data, we provide the measurements here for completeness.

\subsection{Optical Photometry}

The optical measurements in Table 1 were culled from the literature 
(Walker 1969; Warren \& Hesser 1977; Penston 1973; McNamara 1976; 
Rydgren \& Vrba 1984; Penston, Hunter, \& O'Neill 1975; McNamara
et al. 1989). 
Measurements of I magnitudes and V-I colors are available for a subset of 
the sample, largely from the study by Hillenbrand (1997), but 
also from the above references, which have been converted from 
Johnson to Cousins system photometry. 

\subsection{Effective Temperatures and Stellar Luminosities}

Effective temperatures are listed in Table 2 and were derived 
from the spectral types listed in Table 1 and the effective temperature 
scale of Chlebowski \& Garmany (1991) for O-type stars and 
Humphreys \& McElroy (1984) for B0-B3 stars. For stars 
with later spectral types, we used the temperature scales of 
Cohen \& Kuhi (1979) and Bessell (1991). We estimate typical 
uncertainties in log T$_{eff}$ to be 0.05 for stars A5 and earlier; 
0.03 for stars A5 to G5; and 0.02 for stars G5 to K5.

For low mass PMS 
stars of type K and later, veiling emission at V-band can be significant, 
and it is preferable to use I- or J-band measurements to derive
luminosities.  For stars in the current sample, which have much earlier
spectral types, veiling contributions at V-band are negligible.  Therefore,
in order to derive stellar luminosities, we first derive
extinction estimates, 
A$_{v}$ (Table 2), based on spectral types and the observed B-V colors
compared to intrinsic stellar colors (Johnson 1966). 
Next, we derive apparent bolometric luminosities 
from reddening-corrected V magnitudes and bolometric corrections 
tabulated by Code, et al. (1976); Massey, Parker, \& Garmany (1989); 
and Bessell (1991). The bolometric corrections for the PMS stars in our sample 
are all less than 0.25 mag.  We estimate that the standard errors in apparent 
bolometric magnitude are $\pm$0.15 magnitude. 

Conversion of apparent bolometric magnitude to luminosity requires 
a distance modulus. Studies of the Orion association suggest 
that members of the star-forming complex may span distances ranging 
from \ensuremath{\sim}350 to 500 pc (Brown, et al. 1994). Most of the 
stars comprising our sample fall in two well-studied regions 
of the association: Orion Id, a region of size \ensuremath{\sim} 2pc 
centered on the Trapezium cluster, and Orion Ic, also centered 
on the Trapezium but extending 10 pc in projection from \ensuremath{\theta}$^{1}$C. 
The best distance estimates for these sub-regions of the association 
are 480 pc for Orion Id (based on water maser proper
motion and radial velocity measurements; Genzel \& Stutzki 1989) 
and 400 pc for Orion Ic (based on careful photometric study of 
ZAMS B-stars; Warren \& Hesser 1978). Assignment of a sample 
star to the Orion Ic or Id subregions (see column 20 of Table 
1) is based on its \textit{projected} location on the plane of the sky 
in relationship to the Ic/Id boundaries delineated in the map 
published by Warren \& Hesser (1977). This procedure means that 
a modest fraction of the stars assigned as Id could actually 
be Ic stars seen in projection on Id.

The values of log L listed in Table 2 have estimated uncertainties 
of \ensuremath{\pm}0.2 dex, resulting from uncertainties in the assigned 
distances, temperatures, and bolometric corrections. Given values 
of log L and T$_{eff}$, we can derive the stellar radius, which 
is also given in Table 2.

\subsection{Masses, Ages, and Moments of Inertia} 

The luminosities and effective temperatures listed in Table 2 
enable determination of masses, ages, and stellar moments of 
inertia via comparison with PMS evolutionary tracks. The tracks 
published by Swenson et al. (1994; hereafter SFRI) are used here 
because (1) they span the range of masses encompassed by our 
sample; (2) they include not only luminosity (L) and effective temperature
(T$_{eff}$) but also moments of inertia (I) as a function of age; and (3) 
they provide the best matches to open cluster loci 
(Hillenbrand, 2004, in preparation). 
The derived values of mass (M), age (log A), and moment of inertia (log I) 
are listed in Table 2. 
Based on the values of L and T$_{eff}$ and the evolutionary tracks, 
we also specify whether the star is on the convective (C) or 
radiative (R) portion of its evolutionary track, or at the
convective/radiative transition (C/R) in column 9.

Uncertainty in transforming L$_{bol}$ and T$_{eff}$ to mass
and age for PMS stars is dominated by systematic effects 
between different sets of pre-main sequence evolutionary tracks, 
rather than by random errors associated with the luminosity and 
T$_{eff}$ derivations. Unfortunately there are few model-independent 
determinations of masses for stars above the ZAMS (e.g. Mathieu 
et al. 2000). We also note that the SFRI tracks do not take into 
account the effects of accretion during early PMS phases. Accretion 
can dramatically alter both the evolutionary path of PMS stars 
and the radius at which a star of a given final mass joins {\it conventional} 
PMS tracks. In particular, accreting PMS stars follow paths ({\it birthlines}) 
roughly parallel to the ZAMS, at distances above the ZAMS that 
depend on the mass accretion rate (e.g. Palla \& Stahler 1992). 
For example, a young star can in principle reach the same point 
in the HR diagram (a) by following a high accretion rate birthline 
at large radius and then contracting toward smaller radii along 
a {\it conventional} convective track; or (b) by following 
a low accretion rate birthline that deposits it directly at 
relatively low radius. Because of the different possible approaches 
to the main sequence, the ages assigned to non-accreting PMS 
stars and both the ages and masses assigned to accreting PMS 
stars from comparisons of observed luminosities and effective 
temperatures with conventional tracks should be regarded with 
caution. 

\section{THE RELATIONSHIP BETWEEN SPECIFIC ANGULAR MOMENTUM AND MASS
}

\subsection{Overview}

Our measurements of the Orion stars provide for the first time 
a snapshot of the distribution of angular momenta among a sample 
of very young (1-10 Myr) stars, presumably formed under similar 
initial conditions, spanning a mass range from less than 1 M$_{\odot}$ 
to slightly more than 10 M$_{\odot}$, and including a sample of fully 
convective stars with masses in the range 1-2 M$_{\odot}$. 
The mean and median masses within our sample are 2.9 and 2.1 M$_{\odot}$. 
These data allow us to assess systematic trends in stellar angular momenta 
with mass and time.

In this section we first plot our entire sample in the HR diagram 
and describe the systematic trends between surface rotational 
velocity and mass. We then separate the data into two groups: 
1) those stars that are still on their convective tracks and 
provide us with the best estimates of the initial values of specific 
angular momentum (J/M); and 2) more evolved stars that are on 
PMS radiative tracks or on the main sequence.  
We then compare the observationally derived initial values of 
J/M with those of stars at later phases of their evolution and 
identify the processes that most plausibly control the evolution 
of angular momentum.

\subsection{The Data}

Figure 2 summarizes the rotation data for the Orion stars observed 
as part of this survey. We see that the distribution of stars 
in the HR diagram is what we expect for objects that range in 
age from less than one million years to nearly ten million years. 
The more massive stars (M \texttt{>} 4 M$_{\odot}$) are already on the 
main sequence. Intermediate mass stars (2 M$_{\odot}$ \texttt{<} M \texttt{<} 
4 M$_{\odot}$) occupy a variety of positions along the radiative tracks; 
theory (e.g. Palla \& Stahler 1992; Behrend \& Maeder 2001) predicts 
that, depending on the accretion rate, stars with M \texttt{>} 2-4 
M$_{\odot}$ will already be radiatively stable when they reach their 
birthlines and begin quasi-static contraction. Most of the Orion 
stars with M \texttt{<} 2 M$_{\odot}$ are still on their convective tracks.

\subsection{Specific Angular Momentum as a Function of Mass}

The specific angular momentum of a star is given by the relationship
\begin{equation}
J/M = I\omega/M 
\end{equation}
where J is the total angular momentum, M is the mass of the star, 
I is the moment of inertia, and \ensuremath{\omega} is the angular velocity. 
What we measure is \textit{vsini}, and so we can calculate only \textit{Jsini}. 
For a large sample of stars and a random distribution of axes, 
J = (4/$\pi$)J\textit{sini} (for derivation 
see Chandrasekhar \& Munch 1950).

Figure 3a shows the plot of projected specific angular momentum as a function
of mass for stars that are still on their convective tracks along with stars
hotter than log T$_{eff}$ = 4.0, which are already on the ZAMS.  To extend the
data to lower masses, we have also plotted {\it Jsini} values derived from the
{\it vsini} study of low mass ONC stars on convective tracks by Rhode, Herbst, \& Mathieu
(2001).  We have used the SFRI models to derive values of I for each of their
stars.

PMS stars on convective tracks rotate well below the critical 
velocity where centrifugal and gravitational forces balance---the 
so-called breakup velocity. To show this, we have plotted in 
Figure 3a the specific angular momentum that corresponds to rotation 
at the critical velocity for stars on the Palla \& Stahler (1993) 
birthline (hereafter, the PS birthline). The PS birthline intersects 
SFRI convective tracks for masses below 3 M$_{\odot}$ and intersects 
radiative tracks for masses above 3.5 M$_{\odot}$. The discontinuity 
in the breakup velocity curve for stars on the birthline with 
masses between 3 M$_{\odot}$ and 3.5 M$_{\odot}$ comes about because I, 
the moment of inertia, becomes significantly smaller once stars 
join their radiative tracks, at which point they become much 
more centrally condensed. 

Figure 3a shows that there is a fairly smooth and slowly varying 
upper bound to Jsini/M defined by fully convective stars with 
masses less than 2.5 M$_{\odot}$; since these stars are expected to 
rotate as solid bodies, their surface rotation rates should reflect 
true values of J. This upper bound then merges continuously with 
the upper bound determined by higher mass stars that
have already reached the ZAMS.  The five stars that lie
well above the upper bound shown in Figure 3a also lie at the tops
of their convective tracks and are more luminous than most of the
other Orion stars in our sample.  It may be
that, because of depth effects in the cluster, the luminosities and hence
the radii and angular momenta of these five stars are overestimated
or it may be that they are indeed very young and rotating more 
rapidly than the remaining Orion stars (see the discussion in Section 6).

There is about an order 
of magnitude scatter below the upper bound shown in Figure 3a. The
scatter is too 
large to be accounted for by projection effects alone because 
there is only a 14\% chance of observing a rotation rate 
that is less than 50\% of the true equatorial velocity. 
Physical factors that may also contribute to the observed scatter 
are discussed in Section 4.

Kraft (1970) looked at a sample of mature main sequence stars and found 
that a power-law describes the relationship 
between \texttt{<}J/M\texttt{>} and M for {\it main sequence stars} with M \texttt{>} 
2 M$_{\odot}$ (but not those with smaller masses). Specifically, he 
found that \texttt{<}J/M\texttt{>} varies as M$^{0.57}$, and this relationship 
is referred to as the ``Kraft law.'' Kawaler (1987) found the same 
basic relationship but with \texttt{<}J/M\texttt{>} proportional to M$^{1.02}$. 
Kraft and Kawaler both found a marked steepening in the slope
below about 2 M$_\odot$.   
In Figure 3a we see that the power law relationship for PMS stars on
convective tracks extends to masses at least as low as 0.1 M$_{\odot}$.  
The slope of the relationship for PMS convective stars does
{\it not} change 
at 2 M$_{\odot}$ and the value of the slope is about 0.25, somewhat shallower 
than found by Kraft for main sequence stars with M $>$ 2 M$_{\odot}$.  

Figure 3b shows the same plot, but this time for Orion stars 
that are either on their PMS {\it radiative} tracks or on the ZAMS. 
Orion is not old enough to contain stars with M \texttt{<} 1 M$_{\odot}$ 
on radiative tracks. (The data for ZAMS stars with log T$_{eff}$ 
\texttt{>} 4.0 are repeated from Figure 3a.)  We have
42 stars in Orion with M \texttt{>} 3 M$_\odot$, which is a
small number to define the average behavior over a large 
mass range, and these stars 
may not be fully representative, including as they do at least one
He-strong magnetic star (HD 37017) and several other peculiar and
emission line stars.  Therefore, we also show 
in Figure 3b the average values of \texttt{<}Jsini/M\texttt{>} 
for larger sample of main sequence 
field stars as a function of mass. We obtained these values by 
using \texttt{<}\textit{vsini}\texttt{>} data from Abt, Levato,
\& Grosso (2002), 
Abt \& Morrell (1995), and Wolff \& Simon (1997); moments of inertia 
from the SFRI models for ZAMS stars (we extrapolated these models 
to obtain I for masses \texttt{>} 5 M$_{\odot}$); temperatures derived 
from spectral types for the B- and early A-type stars (Cox 2000); 
and masses as a function of temperature from the models of SFRI 
and Maeder \& Meynet (1988). For the F-type stars, we adopted 
the calibrations used by Wolff and Simon.

>From Figure 3b we see that the PMS stars on radiative tracks 
in Orion scatter fairly reasonably
around the mean values for field stars (see also Figure 3c).  There is also a
strong similarity in the overall behavior of the Orion PMS stars
on radiative tracks and the main sequence 
field stars.  Both groups show a relatively slow decline in specific angular
momentum over a mass range of
a factor of 10 (from 30 to 3 M$_\odot$).  From 3 M$_\odot$ to about 1
M$_\odot$ (Orion) or 1.4 M$_\odot$ (field stars), the specific angular
momentum declines by an additional order of magnitude for both groups.
Below about 1.4 M$_\odot$, but not at higher masses, 
stars are believed to lose angular momentum through magnetic
winds during the first 
several hundred million years after they reach the ZAMS (e.g. Kraft 1970; 
Wolff and Simon 1997), and so the slope for M \texttt{<} 1.4 M$_\odot$
depends on the age of the main sequence stars in the sample.
Comparison of PMS stars with main sequence field stars at
 M \texttt{<} 1.4 M$_\odot$ is therefore not relevant. 

These trends in angular momentum as a function of mass are well known 
for main sequence stars.  What is new here is the discovery that
the overall trend of steep decline in average values of angular momentum
for masses between 1 and 2 M$_{\odot}$ {\it is already
present in PMS stars on their radiative tracks and that are
typically no more than a few million years old}.  As Wolff and Simon (1997)
suspected from their study of main sequence stars, an overall decline in
angular momentum along the main sequence from 2 M$_{\odot}$ to
at least 1 M$_{\odot}$ is apparently imposed
during the PMS phase of evolution, and this pattern changes little 
during subsequent main sequence evolution.
This point is illustrated more clearly in Figure 3c, which magnifies the
critical region of Figure 3b.  Here we see that with decreasing mass below 
2 M$_{\odot}$ both the Orion PMS stars on radiative tracks and the field stars 
fall progressively farther away from the power law relationship that described 
the upper bound for PMS convective stars (Figure 3a).

In the sections that follow, we compare the data in Figure 3a on initial 
angular momenta with a very simple model  of star formation to 
illustrate the feasibility of using angular momentum as an additional
constraint on such models. 
We then examine what physical processes might account for the changes in
angular momentum as stars evolve from their convective tracks, as seen
in Figure 3a, onto their radiative tracks, as seen in Figure 3b.

\section{INITIAL ANGULAR MOMENTUM: A CONSTRAINT ON MODELS OF STAR
FORMATION}

The continuity of the upper bound for J/M in Figure 3a suggests 
that the origin of stellar angular momenta is likely similar 
for stars with masses spanning at least two orders of magnitude, 
from 0.1-10 M$_{\odot}$. To illustrate the feasibility of 
relating this observation to theories of star formation, we 
consider the implications of one very simple model, 
which attributes the observed slow
rotation of PMS stars to a magnetic field that is rooted in the 
central star and intercepts the disk (e.g. K\"{o}nigl 
1991).
The net consequence of this star-disk interaction is that 
stellar angular velocity will 
be locked to the Keplerian angular velocity at a co-rotation 
radius until the accretion phase ends and the star is deposited
on the birthline. 
According to K\"{o}nigl, 
the angular velocity of a star locked to its disk is given by:
\begin{equation}
\varpi = \epsilon(GM/R_{in}^{3})^{1/2}
\end{equation}
where \ensuremath{\epsilon} \texttt{<} 1 
is the ratio between the stellar angular 
velocity and the Keplerian velocity at R$_{in}$, i.e. at the radius 
where the disk is disrupted. This radius in turn is given by
\begin{equation}
R_{in} = \beta \mu^{4/7} (2GM)^{-1/7} \mdot^{-2/7}
\end{equation}
where \ensuremath{\beta} is a parameter less than or equal to 1 (with \ensuremath{\beta} 
= 1 corresponding to the classical Alfven radius for spherical 
accretion), \ensuremath{\mu} is the stellar dipole moment, and $\mdot$ is 
the mass accretion rate. Using these equations plus the relationship 
that the surface magnetic field B = \ensuremath{\mu}/R$^{3}$, we then find 
that 
\begin{equation}
J/M = I{\varpi}/M = [I\epsilon(GM)^{5/7}(2)^{3/14} \mdot^{3/7}]/(M\beta^{3/2} B^{6/7} R^{18/7}).
\end{equation}
In order to evaluate this expression, we must make a number of 
assumptions. If we set 
\ensuremath{\epsilon} = \ensuremath{\beta} = 1, then we will 
obtain an upper limit for J/M. We used the SFRI models to obtain 
values of I and R at the birthline. Since these models do not 
extend above 5 M$_{\odot}$, we have not calculated J/M beyond this 
mass limit. We have adopted B = 2500 G, which is typical of the 
limited measurements to date for T Tauri stars (e.g. Guenther, 
et. al. 1999; Johns-Krull, Valenti, \& Koresko 1999).
The number of observational constraints on this assumption is minimal
(magnetic fields have been measured in fewer than 10 T Tauri stars).
However, the observed constancy of magnetic field surrogates, such as
L$_{x}$/L$_{bol}$, over a wide range of PMS star rotation rates (Feigelson et al. 2003)
suggests that this assumption may be acceptable as a first guess.

We have carried out the calculation for 
two birthlines. The first is the PS birthline for 
$\mdot = 10^{-5} M_{\odot} yr^{-1}$ 
(Palla \& Stahler 1993). The second was calculated by Behrend 
\& Maeder (2001), which we will refer to as the BM birthline, 
and is parameterized in terms of luminosity. The BM birthline 
yields results very close to a birthline (Norberg \& Maeder 2000) 
calculated for an accretion rate given by
$\mdot = 10^{-5}\times(1, M^{1.5})$, whichever is larger.

The comparison of the predicted values of J/M with the observed 
Jsini/M values of PMS stars on their convective tracks (i.e. the 
youngest stars in our sample, which are shown in Figure 3a) is 
illustrated in Figure 4. 
It is remarkable that the very simple 
assumptions made here yield both a zero point and a slope that 
are reasonably close to what is observed.
Within the framework of the disk-locking model, 
star-to-star differences in magnetic field strength, 
accretion rates, or in the length of time that disk-locking is 
effective, in addition to projection effects, may all contribute
to the broad scattering of stars below the upper bound. 

Virtually all details of putative
disk-locking mechanisms are currently under debate:  from the linkage of
stellar magnetic fields to the disk to the basic mechanism for angular
momentum loss, i.e. through the disk or through a wind (Shu et al.1994). 
For example,
within the past year, Johns-Krull \& Gafford (2002) have extended 
the Shu et al. model to include complex, non-dipole field topologies, 
which they argue provide a more realistic representation of currently 
available observations. 
While this model has many attractive features, 
it and other more complex formulations introduce additional parameters,
which must be derived empirically and which have not yet been well 
established.  Nevertheless, the approximate coincidence between
the observed J/M vs. M relationship and the predictions of this very
simple model should serve as a challenge to theorists to develop
a truly predictive
theory of angular momentum loss during the disk accretion phase.

\section{FROM THE BIRTHLINE TO THE MAIN SEQUENCE}

Next we turn to the question of what happens to angular momentum 
as intermediate-mass PMS stars evolve from convective to radiative tracks 
and finally onto 
the main sequence. The observable diagnostic is the surface rotational 
velocity, which in turn depends on the initial angular momentum 
plus any modifications caused either by the transport of angular 
momentum from the stellar interior or by external forces, including 
mass accretion and mass loss, particularly when the 
stellar magnetic field is strong. In the sections that follow, 
we first look for evidence relating to the question of whether 
or not external forces are likely to alter the initial stellar 
angular momentum of pre-main-sequence stars. We then look at 
the issue of how structural changes affect the observed rotational 
velocities. Informed by this discussion, we predict ZAMS rotational 
velocities for the stars in our sample, and compare these predictions 
with observations of stellar rotation among young field stars. 
This provides the basis for evaluating how well trends of rotation 
with mass along the main sequence can be predicted from patterns 
already present during early PMS evolution. 

\subsection{External Forces: Disk Regulation on Convective Tracks}

If no external forces were acting, pre-main-sequence stars should 
spin up as they contract toward the zero-age main sequence because 
of structural changes: a decrease in radius and an increase in 
central concentration.  Some stars may also  
lose angular momentum as they evolve down their convective 
tracks, possibly as a consequence of magnetic torques that transfer 
angular momentum away from the star to a surrounding disk. 
Our current sample does not allow us to search for systematic 
changes in J/M as stars with M \texttt{>} 1 M$_{\odot}$ evolve down their 
convective tracks. The reasons are two: 1) our sample spans a 
relatively large mass range but has relatively few stars at any 
given mass that are still on their convective tracks, and 
we cannot subdivide the sample according to mass and position 
along the convective track and retain sufficient numbers to average 
out star-to-star fluctuations in \textit{v} and \textit{sini}; and 2) these 
relatively massive stars evolve down only a truncated portion 
of the convective track (Palla \& Stahler 1992) and do not, therefore, 
change in radius by more than about a factor of 2 ; as a result, 
any systematic changes in \textit{vsini} will be small relative to 
the dispersion in \textit{v} at a given position along the convective 
tracks.  As we discuss in Section 6, however, the systematic 
differences in J/M for stars on convective and radiative tracks
do appear to require some loss of angular momentum as stars
with M \texttt{<} 2 M$_{\odot}$
evolve down their convective tracks.

\subsection{External Forces: Mass Loss}

Stellar winds loaded onto open magnetic field lines can exert 
a spindown torque on stars. This mechanism has, however, been 
shown to be ineffective for fully convective stars because the 
PMS timescale for spindown exceeds the evolutionary timescale 
by a few orders of magnitude (e. g. MacGregor \& Charbonneau 
1994). Fully convective stars are assumed to rotate as solid 
bodies, and the wind must slow down the entire star---which fact 
precludes significant spindown during the relatively short PMS convective 
phase. 

Mass loss is also unlikely to have a significant effect on the 
angular momentum of most stars with M \texttt{>} 1 M$_{\odot}$ 
that are evolving 
along their radiative tracks.  The majority of these stars are hotter 
than log T$_{eff}$ = 3.8, do not have surface convective zones, and 
are not expected to have dynamo-generated magnetic fields. This 
expectation is borne out by observations of x-rays from PMS stars 
in Orion. X-ray emission can serve as a proxy indicator of magnetic 
fields. Observations show that the ratio of x-ray to bolometric 
luminosity drops by more than two orders of magnitude when PMS 
stars become fully radiative (Flaccomio, et al. 2003), and
x-ray observations of PMS Orion stars show that strong x-ray 
emission is present only in stars cooler than log T$_{eff}$ = 3.8 
(Gagne, et al. 1995; Feigelson et al 2002).  About 10 percent 
of the stars with spectral types around A0 and earlier,  
including the He-rich star HD 37017 in our sample, do have 
large (several thousand gauss) magnetic fields which are
thought to be fossil rather than
dynamo-generated fields, and it is not known how these large
fields might affect either the initial angular momentum or the
subsequent evolution of angular momentum.  However,
the number of such peculiar stars is small enough that,
even if some are included 
in the sample, they should not affect the overall trends shown
in Figure 3.

For the purposes of this paper, we will assume very few
stars hotter than log T$_{eff}$ = 3.8 have strong surface fields,
and they are therefore unlikely to experience strong braking torques 
caused by stellar winds.

\subsection{Internal Effects: Changes in the Moment of Inertia}

Even if external forces play no role, we do expect that changes 
in the moment of inertia will affect stellar rotation. As a star 
contracts and moves down its convective track and then along 
a radiative track, its radius decreases and its central concentration 
increases. Both effects should increase the rotation rate. Whether 
or not increasing central concentration has an effect on the 
rotation rate at the {\it stellar surface} depends on the time scale 
for the transport of angular momentum from the stellar core to 
the surface of the star. Since the theory of this process has 
substantial uncertainties, we will bracket the true situation 
by considering two extreme cases for how angular momentum might 
be conserved: 1) conservation of angular momentum in shells; 
and 2) solid body rotation.

If angular momentum cannot be efficiently transported across 
adjacent spherical shells or from the radiative core to the convective 
envelope, then at any given time \textit{t} the rotational velocity \textit{v} 
is given by 
\begin{equation}
v(t) = v_0 \times R_0/R(t),
\end{equation}
where R is the radius of the star and v$_{0}$ and R$_{0}$ refer to initial 
values.

If, on the other hand, angular momentum is transported efficiently 
throughout the star, and the star effectively rotates as a solid body, then
\begin{equation}
v(t) = v_0 \times (I_0/R_0) \times R(t)/I(t),
\end{equation}
where I is the stellar moment of inertia.

Figure 5 compares these two cases for four different stellar 
masses, based on the models by SFRI. We have assumed arbitrarily 
a rotation of 10 km/s for the starting point on the tracks. Three 
different regions should be distinguished in this figure: 1) 
the fully convective phase, which we will take to be the evolution 
that takes place at nearly constant T$_{eff}$ to the point of minimum 
luminosity; 2) the transition from the convective to the radiative 
tracks, which we will take to be the portion of the evolutionary 
tracks from minimum luminosity to log T$_{eff}$ = 3.8, at which 
temperature all of the stars in our sample are on radiative tracks; 
and 3) the remainder of the radiative track to the main sequence.

As Figure 5 shows, during the fully convective phase, the two 
extreme cases for angular momentum conservation predict essentially 
the same evolution of surface rotation rate. As stars evolve 
down their convective tracks, contraction is nearly homologous, 
and changes in \textit{I} directly track changes in \textit{R}$^2$. Along 
the radiative tracks, the two extreme cases yield nearly parallel 
tracks in this logarithmic plot (or for the lowest mass case, 
predict a relatively small change in surface rotation). Therefore 
observations of samples comprising either fully radiative or 
fully convective stars alone cannot be used to determine empirically 
how angular momentum is conserved. In fact, either assumption
(solid body rotation or conservation of angular momentum in shells)
can be used to predict how the surface rotation will change either
\textit{along} radiative 
or \textit{along} convective tracks. Because of the simplicity of 
the calculation, we will assume that angular momentum varies 
inversely with radius along both convective and radiative tracks.

The situation is very different during the \textit{transition} from 
convective to radiative tracks. It is at this point in its evolution 
that a star begins to develop a highly concentrated radiative 
core, and this core can in principle store much of the initial 
stellar angular momentum. Therefore, it is during this transition 
that major differences develop in the predicted
\textit {surface rotation} 
rates depending on how angular momentum is conserved. If the 
star were to rotate as a solid body during this phase of evolution, 
the rotation rate would increase by a factor of 4 or more relative 
to the rate observed on the convective track (cf. Figure 5).

The observed increase is, however, less than a factor of 4. In 
our own sample, there are 21 stars on convective tracks with 
masses greater than 1.2 M$_{\odot}$, and for these stars \texttt{<}\textit{vsini}\texttt{>} 
= 32 $\pm$ 3 km/s. There are 11 stars in this mass range with 
3.775 \texttt{<} log T$_{eff}$ \texttt{<} 3.826, i.e. at the beginning of 
their radiative tracks. For these 11 stars, \texttt{<}\textit{vsini}\texttt{>} 
= 65 $\pm$ 14 km/s. The observed spin up is only a factor of 
two or so, much smaller than predicted for solid body rotation. 
Most of the convective stars have log T$_{eff}$ in the range 3.65-3.70, 
and reference to Figure 5 shows that, if angular momentum is
conserved in shells, the spinup as stars evolve 
from this temperature range to log T$_{eff}$ = \ensuremath{\sim}3.8 is about 
a factor of two for stars with M \texttt{>} 1.2 M$_{\odot}$
in agreement with the observations.

Therefore, we conclude that ``core-envelope decoupling'' occurs 
as PMS stars evolve from convective to radiative tracks. That is, 
stars that start their post-birthline evolution along convective 
tracks develop rapidly rotating radiative cores and slowly rotating 
envelopes as they transition from PMS convective to radiative 
tracks. A number of studies of stars with mass of a solar mass 
and lower have also concluded that core-envelope decoupling must 
occur at some point during PMS evolution. These papers have addressed 
the question of why there are a large number of slowly rotating 
stars with masses near 1 M$_{\odot}$ in young clusters and have argued 
that core-envelope decoupling provides the best explanation (e.g. 
Krishnamurthi, et al. 1997; Allain 1998; Barnes, Sofia, \& Pinsonneault 
2001; and Soderblom, Jones, \& Fischer 2001). Our observations 
of PMS stars that span the transition from convective to radiative 
phases provide direct confirmation of these inferences, but as we
shall see in the next section, core-envelope decoupling accounts for
only part of the loss of angular momentum that occurs between the
birthline and the main sequence.

\section{COMPARISON OF PREDICTED AND OBSERVED ZAMS ROTATIONAL VELOCITIES
}

\subsection{Stars on PMS Convective Tracks}

The preceding discussion argues that stars evolving from convective 
to radiative PMS tracks develop rapidly rotating cores and decoupled, 
slowly rotating envelopes. We expect that changes in observed 
surface rotation will therefore track the changes in stellar 
radius. We can check this conclusion by predicting the rotation 
rates that stars in our sample on convective tracks will have 
when they reach the ZAMS. The average mass of the stars on convective 
tracks in our sample is 1.74 M$_{\odot}$. A good comparison sample 
is the group of stars closest to the ZAMS with masses in the 
range 1.5-2 M$_{\odot}$ studied by Wolff \& Simon (1997). In order 
to predict the rotation rates that the PMS Orion stars will have 
when they reach the ZAMS, we have used SFRI models to obtain 
both the ZAMS radii and the moments of inertia as a function 
of mass. Figure 6 shows the results, where we have predicted 
the ZAMS rotation rates for conservation of angular momentum 
in shells (that is, assuming core-envelope decoupling), and for 
solid body rotation. We see that the assumption of solid body 
rotation produces a ZAMS distribution that has, as anticipated 
from the above discussion, essentially no overlap with what is 
observed for field stars. There is, however, good agreement between 
the predicted and observed distributions if we assume that the 
rotation rate varies inversely with stellar radius as expected 
for core-envelope decoupling.

\subsection{Stars on Radiative Tracks}

We have identified no mechanisms that would cause loss of
angular momentum after stars reach log T$_{eff}$ = 3.8 as they
evolve toward hotter temperatures and smaller radii.
These stars should therefore conserve angular momentum as
they complete their evolution to the main sequence.  In
order to check this prediction, 
we will take as our initial condition the range of \textit{vsini} 
values observed for stars with temperatures in the range log 
T$_{eff}$ = 3.75-3.85. Note that this temperature
range encompasses the region of the HR diagram within which stars 
of masses 1.5-3.5 M$_{\odot}$ transition from convective to radiative 
tracks. The data in Table 2 show that the Orion stars reach this 
temperature with a range of rotation rates from \texttt{<}20 km/s 
to nearly 200 km/s, with the majority of stars having rotations 
less than 150 km/s (see Figure 7).
In accord with our previous discussion, we will assume that the 
rotation rates of these stars then vary inversely with the
radius (see also Hartmann, et al. 1986).

In Figure 8 we have plotted the predicted changes in \textit{vsini} 
as stars evolve along radiative tracks to the main sequence. 
This figure illustrates several points. First, the rotational 
velocities of most of the stars in the current sample lie between 
the bands defined by evolving models for initial values of \textit{vsini} 
equal to 20 and 150 km/s. Therefore the \textit{range} of rotational 
velocities observed at the end point of the contraction, namely 
along the ZAMS, is fully consistent with the \textit{range} seen
among pre-main-sequence 
stars. Hence, as expected, there is no evidence for significant 
loss of angular momentum along radiative tracks.

Second, on the basis of these simple models we would expect \textit{vsini} 
on average to increase with increasing mass because more
massive stars contract 
more as they traverse their longer radiative tracks, and this
expectation is consistent with the observations up to about
3 M$_{\odot}$. 

Figure 8 shows the prediction that rotation will continue to increase 
to values in excess of 400 km/s for stars with masses greater 
than 4 M$_{\odot}$ is not borne out by the observations. 
While our sample is 
small, this conclusion is supported by much larger surveys (e.g 
Abt, Levato, \& Grosso 2002; Wolff, Edwards, \& Preston 1982). 
This result has a natural explanation in terms of the birthline. 
Stars with masses greater than 3.5 M$_{\odot}$, given our particular 
choice of models, are already on radiative tracks when accretion 
stops.  The portion of the radiative track that they traverse
decreases with increasing mass, and the corresponding spin-up
from the initial conditions when they are released at the birthline
will also decrease, thereby limiting the maximum observed
rotational velocity.  

\subsection{The Break in the Power Law}

We now turn to an explanation of why J/M decreases sharply with M
for stars with masses less than about 2 M$_{\odot}$ that have 
completed the convective phase of evolution. 
We have noted already that core-envelope decoupling during 
the transition from convective to radiative tracks will account 
for some apparent loss of observed surface angular momentum. 
Core-envelope decoupling cannot, however, account for the sharp
downturn in J/M.  
The effects of core-envelope decoupling are predicted 
to be largest for stars around 2 M$_{\odot}$ and to diminish toward 
lower masses (see Figure 5). This prediction is in the opposite
sense to the trend seen in the observations:  the differences between
the angular momenta of stars on convective tracks
and the values for stars on radiative tracks (compare Figures 3a
and 3b) are much larger at 1 M$_{\odot}$ than at 2 M$_{\odot}$.

An additional mechanism is apparently required to explain 
the observations, and we
suggest that this mechanism is braking while stars 
with M \texttt{<} 2 M$_{\odot}$ evolve down their convective tracks. 
Stars on convective tracks have accretion rates on the order
of 10$^{-8}$ M$_{\odot}$ per year (e. g. 
Valenti, Basri, \& Johns 1993; Gullbring et al. 1998).  If these stars
are locked to their disks, then the 
disks could in principle act as a brake and cause additional loss of angular
momentum. 

Hartmann (2002) has pointed out that three factors will determine the amount of spindown for a PMS 
star: 1) the time spent on the convective track; 2) the time 
scale for disk-braking; and 3) the lifetime of the disk.
He has estimated that the disk-braking time scale 
is given by
\begin{equation}
\tau_{DB} = 4.5 \times 10^6 {\rm yr} [{M\over{0.5 M_\odot}}] [{10^{-8} M_\odot {\rm yr}^{-1} \over{\mdot}}] f,
\end{equation}
where the mass and accretion rate parameters are scaled to values
typical of low mass T Tauri stars (Gullbring et al. 1998; 
Hartmann et al. 1998), and f is the ratio of the actual velocity 
to the breakup velocity. We will take, following Hartmann (see 
also Figure 3a), f = 0.2. We can then calculate the ratio of 
$\tau_{DB}$ to the time a contracting star spends moving down the 
convective track from the birthline to the transition to the 
radiative track, after which we expect braking to be minimal because 
of the absence of magnetic fields. In order to estimate the convective 
lifetime, we have used the SFRI models to determine the time 
that elapses as a star moves from the PS birthline to the bottom 
of the convective track.

The results are shown in Figure 9. The shape of this curve mimics 
the shape of the relationship between J/M and mass for PMS Orion 
stars on radiative tracks. 
The lower the mass of a star, the longer the time it 
spends evolving down its convective track and the more angular 
momentum it can lose, provided of course that the disk lifetime
is also long enough.  Stars with masses close to 2 M$_{\odot}$ 
enter the radiative phase before disks can remove significant 
stellar angular momentum.  Therefore braking on the convective 
track provides a very natural explanation for the break in 
the power law seen for post-convective-track stars with M \texttt{<} 
2 M$_{\odot}$.  Analysis of 
data for Orion and several hundred other PMS stars with masses 
in the range 0.5-1 M$_{\odot}$ shows that these stars on average reduce 
their rotational velocities by about a factor of three while 
contracting by about a factor of three as they evolve down convective 
tracks, thereby reducing their angular momentum by about an order 
of magnitude (Rebull et al. 2002; Rebull et al. 2003, in preparation). 
Quantitatively, this is just what is required in order to account 
for the break in the J/M power law (see Figure 3b).

\section{CONCLUSIONS}

We have presented new measurements of rotational velocities for a sample
of stars with masses in the range 0.4-14 
M$_\odot$ (median mass 2.1 M$_{\odot}$) 
and considered literature data for stars $<$0.5 M$_\odot$. 
Observations 
of the youngest stars in Orion show that the specific angular 
momentum of stars on \textit{convective} tracks increases slowly and 
continuously with stellar mass over the mass range from 0.1 M$_{\odot}$ 
to nearly 3 M$_{\odot}$ and merges smoothly with the J/M vs. M relationship 
for young main sequence stars with masses between 3 and 10 M$_{\odot}$. 
The power law relationship between J/M and M for newly formed 
stars suggests a common mechanism for establishing J/M throughout 
this entire range, which spans a factor of 100 in mass. 
The power law relationship between J/M and M for convective PMS 
stars differs significantly from what is observed for both older PMS 
stars, which have evolved from convective to radiative tracks,
and main sequence stars. For these 
older stars, J/M follows the same power law relationship as the convective
stars for M \texttt{>} 2 M$_{\odot}$ but decreases sharply 
with decreasing mass for stars 
with M \texttt{<} 2 M$_{\odot}$. 

These observations establish the basic trends in angular momentum
as a function of time and mass that models must explain.  Comparison
with very simple models shows that these overall trends
can be explained by five distinct processes that are effective at
different stages of evolution:  1) an angular momentum loss process,
possibly disk-locking, that operates before 
stars reach the birthline and applies to 
all stars with masses between 0.1 and 10 M$_{\odot}$; 2) braking of 
stars with M \texttt{<} 2 M$_{\odot}$ as they
evolve down their convective tracks, with the amount of braking
increasing with time spent in this phase of evolution and hence with
decreasing mass; 3) decoupling of
the angular momentum seen at the surface of the star from the angular
momentum in the interior when stars with M \texttt{<} 2-4 M$_{\odot}$
make the transition
from convective to radiative PMS evolution; 4) conservation of angular
momentum as stars evolve along their radiative tracks; and 5) additional
spindown by magnetic winds of stars with M \texttt{<} 1.4 M$_{\odot}$
after these stars reach the main sequence. 

While much more sophisticated models may eventually be required to
explain the trends observed here, 
these results show the potential power of observations of stellar
rotation as probes of fundamental processes that occur during the formation and 
early evolution of stars.





\tablenum{1}
\pagestyle{empty}

\begin{deluxetable}{c r r r c r r c r r r c c c c c r r r c}
\rotate
\tabletypesize{\scriptsize}
\tablewidth{675pt}
\tablecaption{Observed Parameters} 
\tablehead{
\colhead{ Par. }&
\colhead{ V  }&
\colhead{ B-V }&
\colhead{ U-B }&
\colhead{ UBV \tablenotemark{a}}&
\colhead{ Ic }&
\colhead{ V-Ic}&
\colhead{ VI \tablenotemark{b}}&
\colhead{ J  }&
\colhead{ J-H }&
\colhead{ H-K  }&
\colhead{ JHK \tablenotemark{c}}&
\colhead{ SpT   }&
\colhead{ SpT \tablenotemark{d} }&
\colhead{ vsini \tablenotemark{e} }&
\colhead{ vsini \tablenotemark{f}}&
\colhead{ Brun }&
\colhead{ HD  }&
\colhead{ HR }&
\colhead{ Group} \\
      \colhead{ }&
      \colhead{ mag }&
      \colhead{ mag }&
      \colhead{ mag } &
      \colhead{ ref } &
      \colhead{ mag }&
      \colhead{ mag }&
      \colhead{ ref}&
      \colhead{ mag} &
      \colhead{ mag} &
      \colhead{ mag} & 
      \colhead{ ref}&
      \colhead{ } &
      \colhead{ ref  }&
      \colhead{ km/s }&
      \colhead{ ref   }  &
      \colhead{ \#    }&
      \colhead{ \#    } &
      \colhead{ \#  } &
      \colhead{ }
}
\startdata

    82 &  7.97 & -0.04 &-0.35 &    2 & 7.98 &-0.01 &        -- &\nodata &\nodata &\nodata&        -- & B8-B9,B9 & 10,11&       200 &     8&     -- & 36120&     -- &     C  \\ 
   378 &  8.65 & -0.06 &-0.38 &    2 & 8.75 &-0.10 &        -- & 8.79 &-0.03 &-0.04 &        1 & B8-A0,A0 & 10,11&        15 &     8&     -- & 36234&     -- &     C  \\ 
   597 &  8.18 &  0.05 &-0.04 &    2 & 8.07 & 0.11 &        -- &\nodata &\nodata &\nodata&        -- &       B9 & 10,11&        25 &     8&     -- & 36366&     -- &     C  \\ 
   679 &  6.22 & -0.18 &-0.75 &    2 &\nodata &\nodata&        -- & 6.66 &-0.06 &-0.07 &        1 &  B2.5,B2 & 12,7,11&  15,10,40 & 4,10,12&     -- & 36430&   1848 &     C  \\ 
   854 &  7.69 & -0.08 &-0.45 &     2 & 7.74 &-0.05 &        -- & 7.88 &-0.03 &-0.04 &        1 &  B5.5,B8 & 12,11&       190 &     8&     -- & 36541&     -- &     C  \\ 
   908 &  8.81 & -0.04 &-0.22 &    2 & 8.85 &-0.04 &        -- & 8.89 &-0.01 &-0.04 &        1 &    B8-B9 & 10,11&       135 &     8&     -- & 36559&     -- &     C  \\ 
  1044 &  7.69 &  0.01 &-0.65 &    1 &\nodata &\nodata&        -- & 7.58 & 0.01 &-0.01 &        1 &    B2-B3 & 1,3,11&       $<$50 &     1&    25  & 36629&     -- &    C1  \\ 
  1049 & 11.87 &  1.60 & 1.54 &    1 &\nodata &\nodata&        -- & 8.78 & 0.05 & 0.89 &        1 &    K5-K7 &     1&  4-20 (SB:50) &     1&     29 &    --&     -- &    C1  \\ 
  1076 & 12.60 &  0.47 &\nodata &    5 &\nodata&\nodata&        -- &10.26 & 0.61 & 0.45 &        1 &       K1 &     9&        22 &    11&     42 &    --&     -- &     C  \\ 
  1097 &  8.63 & -0.04 &-0.08 &    1 &\nodata &\nodata&        -- & 8.62 & 0.06 & 0.00 &        1 &    B8,B9 &  3,11&       150 &     8&     50 & 36655&     -- &     C  \\ 
  1126 &  8.98 &  0.01 &-0.06 &    1 &\nodata &\nodata&        -- & 8.97 & 0.00 &-0.03 &        1 &    B9-A0 & 1,3,11&        80 &     1&     62 & 36670&     -- &    C1  \\ 
  1179 & 11.52 &  0.53 &-0.02 &    1 &\nodata &\nodata&        -- &10.42 & 0.24 & 0.03 &        1 &    F5,F8 &   1,3&       114 &     1&     89 &    --&     -- &    C1  \\ 
  1270 & 12.11 &  0.92 & 0.68 &    1 &11.01 & 1.10 &        4 &10.28 & 0.58 & 0.27 &        1 &       K1 &     9&        16 &    11&    141 &    --&     -- &     C  \\ 
  1319 & 12.58 &  1.11 & 0.53 &    1 &\nodata &\nodata&        -- &10.29 & 0.54 & 0.15 &        1 & F5-G0,K0 &   1,4&   187,279 &   1,6&    166 &    --&     -- &    C3  \\ 
  1322 & 11.70 &  0.85 & 0.33 &    1 &10.76 & 0.94 &        4 &10.12 & 0.40 & 0.10 &        1 &       G8 &     1&        70 &     1&    168 &    --&     -- &    C2  \\ 
  1326 & 11.34 &  1.46 & 1.31 &    2 &\nodata &\nodata&        -- & 8.45 & 0.64 & 0.17 &        1 &       K5 &     1&  4-20 (SB:13) &     1&    172 &    --&     -- &    C1  \\ 
  1345 & 12.00 &  1.26 &\nodata &    5 &\nodata&\nodata&        -- & 9.19 & 0.70 & 0.17 &        1 &       K5 &     1&      4-20 &     1&     -- &    --&     -- &    C1  \\ 
  1360 & 13.81 &  0.94 & 0.42 &    1 &\nodata &\nodata&        -- &\nodata &\nodata&\nodata&        -- &       G8 &     4&        15 &     3&    182 &    --&     -- &    C2  \\ 
  1374 & 10.31 &  0.56 & 0.10 &    4 & 9.66 & 0.65 &        4 & 9.14 & 0.27 & 0.04 &        1 &       F6 &     1&        77 &     1&    203 &    --&     -- &    D1  \\ 
  1391 & 10.63 &  0.49 & 0.04 &    1 &10.11 & 0.51 &        2 & 9.68 & 0.23 & 0.06 &        1 &       F7 &   1,3&  15 (SB:30) &     1&    211 &    --&     -- &    C3  \\ 
  1393 & 12.16 &  0.78 & 0.37 &    4 &11.08 & 1.04 &        1 &10.32 & 0.53 & 0.08 &        2 &    K0,G6 &   1,2&        10 &   3,1&    213 &    --&     -- &    D1  \\ 
  1394 & 10.29 &  0.55 &-0.01 &    1 & 9.44 & 0.69 &        2 & 9.00 & 0.37 & 0.30 &        1 &       F6 &   1,3&        60 &     1&   216  &    --&     -- &    D1  \\ 
  1404 & 11.51 &  0.84 & 0.34 &    1 &10.56 & 0.95 &        5 & 9.81 & 0.60 & 0.47 &        1 &       G5 &   3,5&  34,27,14 & 3,11,7&    220 &    --&     -- &    D1  \\ 
  1408 & 14.40 & -0.75 &\nodata &    5 &\nodata&\nodata&        -- &10.45 & 0.59 & 0.14 &        1 &    K5-K7 &     1&        21 &     1&    219 &    --&     -- &    C1  \\ 
  1409 & 11.60 &  0.85 & 0.28 &    1 &10.59 & 1.01 &        5 & 9.72 & 0.69 & 0.57 &        1 &    G4:F8 &   1,4&  29,45,60 & 1,5,6&    224 &    --&     -- &    C3  \\ 
  1414 & 11.48 &  0.64 & 0.11 &    1 &10.72 & 0.70 &        5 &10.27 & 0.28 & 0.03 &        1 &    F8,G5 &   1,3&  29,38,14 & 1,3,7&    225 &    --&     -- &    C1  \\ 
  1425 & 12.00 &  0.91 & 0.34 &    1 &11.01 & 1.00 &        4 &10.15 & 0.41 & 0.15 &        1 &       G7 &     1&     26,33 &   1,3&    233 &    --&     -- &    C1  \\ 
  1440 & 12.73 &  0.96 & 0.49 &    1 &\nodata &\nodata&        -- &11.17 & 0.48 & 0.07 &        1 & K5-7,K0-2 &   1,4&     16-20 &     1&   244  &    --&     -- &    C2  \\ 
  1445 &  8.15 & -0.09 &-0.45 &    1 &\nodata &\nodata&        -- & 8.33 &-0.02 &-0.03 &        1 &    B6,B7 & 1,3,11&   280,245 &   1,8&    246 & 36842&     -- &    C1  \\ 
  1455 & 10.89 &  0.64 &-0.16 &    4 &10.08 & 0.76 &        1 & 9.56 & 0.35 & 0.11 &        1 &       G0 &   1,2&        21 &     1&    252 &    --&     -- &    D1  \\ 
  1484 & 12.20 &  1.08 &\nodata &    5 &10.97 & 1.04 &        1 &10.27 & 0.49 & 0.11 &        1 &       K1 &     1&     49,44 &   1,3&    283 &    --&     -- &    D1  \\ 
  1491 &  7.44 & -0.06 &-0.39 &    1 &\nodata &\nodata&        -- & 7.54 & 0.01 & 0.00 &        1 &    B8-B9 & 1,3,11&   225,220 &   1,8&    281 & 36865&     -- &    C1  \\ 
  1505 & 12.78 &  0.87 & 0.48 &    1 &\nodata &\nodata&        -- &11.06 & 0.57 & 0.09 &        1 &      K2: &     1&  17-20,30 &   1,3&    293 &    --&     -- &    C1  \\ 
  1507 & 10.26 &  0.28 & 0.21 &    1 & 9.96 & 0.30 &        4 & 9.70 & 0.10 & 0.03 &        1 & A5,A7,A8 & 1,3,11&   136,235 &   1,9&    295 & 294265&     -- &    C3  \\ 
  1510 & 12.40 &  1.25 &\nodata &    5 &11.09 & 1.31 &        1 &10.08 & 0.65 & 0.19 &        2 &       K2 &     1&    38,39  & 1,14 &    302 &    --&     -- &    D1  \\ 
  1511 &  9.30 &  0.15 & 0.15 &    1 & 9.14 & 0.16 &        4 & 9.05 & 0.04 &-0.01 &        1 &       A2 & 1,3,11&   112,180 &   1,9&    304 & 36866&     -- &    D1  \\ 
  1518 & 13.80 &  0.80 &\nodata &    9 &11.65 & 1.60 &        1 &10.28 & 0.72 & 0.24 &        2 &       K2 &     2&        38 &     3&    311 &    --&     -- &    D1  \\ 
  1539 & 10.77 &  0.71 & 0.29 &    1 & 9.80 & 0.94 &        1 & 9.10 & 0.17 & 0.09 &        2 &       B8 &   1,3&       250 &     1&    328 &    --&     -- &    C3  \\ 
  1540 & 11.35 &  1.27 & 1.11 &    3 & 9.89 & 1.49 &        1 & 8.83 & 0.55 & 0.23 &        2 &    K4,K1 &   1,2&  15-20 (SB) &     1&    334 &    --&     -- &    D1  \\ 
  1541 & 12.58 &  0.99 & 0.41 &    1 &11.12 & 1.27 &        1 &10.14 & 0.53 & 0.17 &        2 & K1,K2,K3 & 1,5,4&  11-20,29,15 &   1,3,14&    335 &    --&     -- &    D1  \\ 
  1552 & 13.72 &  1.00 &-0.65 &    8 &11.65 & 1.75 &        1 &10.07 & 0.98 & 0.71 &        2 &       K7 &     2&        15 &     3&    340 &    --&     -- &    D1  \\ 
  1553 & 12.40 &  1.36 &\nodata &    5 &\nodata&\nodata&        -- &10.09 & 0.77 & 0.47 &        1 &       K2 &     1&  28 (SB:54) &     1&    339 &    --&     -- &    D1  \\ 
  1554 & 12.30 &  0.88 & 0.43 &    1 &\nodata &\nodata&        -- &10.75 & 0.50 & 0.09 &        1 &    G8,K2 &   1,4&        30 &     1&   341  &    --&     -- &    C4  \\ 
  1562 &  9.47 &  0.08 & 0.03 &    1 & 9.59 & 0.05 &        3 & 9.51 &-0.01 &-0.03 &        1 &    A0,B9 & 1,3,11&   200,250 &   1,9&   342  & 36899&     -- &    C3  \\ 
  1581 & 12.46 &  0.73 & 0.27 &    1 &\nodata &\nodata&        -- &10.76 & 0.34 & 0.09 &        1 &       F5 &     1&  17-20 (SB:12) &   1,3&   365  &    --&     -- &    C1  \\ 
  1587 & 12.60 &  1.30 &\nodata &    5 &11.11 & 1.61 &        1 & 9.84 & 0.63 & 0.21 &        2 &       K2 &   1,2& 19-20,32  & 1,14 &    374 &    --&     -- &     D  \\ 
  1605 &  8.03 &  0.17 & 0.10 &    1 & 7.57 & 0.41 &        1 & 7.27 & 0.28 & 0.51 &        1 &    A0,B9 & 1,11,2&  127,170,110 & 1,9,8&   388  & 36917&     -- &    D1  \\ 
  1608 & 12.95 &  1.03 & 0.79 &    1 &\nodata &\nodata&        -- &10.91 & 0.63 & 0.10 &        1 &    K5,K4 &   1,4&        26 &     1&    394 &    --&     -- &    C4  \\ 
  1623 & 10.13 &  0.57 & 0.43 &    4 & 9.29 & 0.90 &        1 & 8.46 & 0.48 & 0.49 &        1 &    A3,A2 &   1,2&   171,260 &   1,9&    405 &    --&     -- &    C3  \\ 
  1626 & 11.24 &  0.68 & 0.14 &    1 &10.48 & 0.77 &        5 & 9.89 & 0.31 & 0.07 &        1 &    A5:G2 &   1,6&    94,110 &   1,7&   407  &    --&     -- &    C4  \\ 
  1634 &  8.39 & -0.10 &-0.48 &    1 &\nodata &\nodata&        -- & 8.93 &-0.06 &-0.04 &        1 &  B7,B9IV & 1,3,11&       150 &     1&    417 & 36918&     -- &    C4  \\ 
  1643 & 13.30 &  0.63 &\nodata &    5 &\nodata&\nodata&        -- &10.95 & 0.54 & 0.18 &        1 &       K0 &     1&  16-20,36 &   1,3&    424 &    --&     -- &    D1  \\ 
  1646 &  9.80 &  0.49 & 0.01 &    1 &\nodata &\nodata&        -- & 7.81 &-0.03 &-0.03 &        1 &       F6 &   1,3&     11-20 &     1&    425 & 294257&     -- &    C1  \\ 
  1654 &  8.88 &  0.08 &-0.15 &    1 &\nodata &\nodata&        -- & 8.87 & 0.21 & 0.03 &        1 &    B8-B9 & 1,3,11&       120 &     1&    437 & 36938&     -- &    C2  \\ 
  1657 & 11.51 &  0.51 &-0.03 &    1 &10.93 & 0.58 &        4 & 8.58 & 0.06 & 0.03 &        1 &    F2,G2 &   1,4&     14-20 &     1&    438 &    --&     -- &    C4  \\ 
  1659 & 11.62 &  1.21 & 0.77 &    3 &10.15 & 1.56 &        1 & 9.03 & 0.64 & 0.30 &        2 &    K2,K3 &   1,2&  9-20,7 (SB:30) & 1,14 &    443 &    --&     -- &     D  \\ 
  1660 &  9.00 & -0.02 &-0.22 &    1 & 8.93 & 0.00 &        1 & 8.93 & 0.14 & 0.00 &        2 &    B8,B9 & 1,11,2&   275,375 &   1,9&    442 & 36939&     -- &    D1  \\ 
  1664 &  7.59 & -0.13 &-0.55 &    1 &\nodata &\nodata&        -- & 8.92 & 0.05 & 0.03 &        1 &    B3,B5 & 1,3,11&      180: &     1&   440  & 36936&     -- &    C1  \\ 
  1671 &  9.65 &  0.26 & 0.11 &    1 &\nodata &\nodata&        -- & 9.06 & 0.13 & 0.00 &        1 &    A5-A7 & 1,3,11&       175 &     1&    454 & 36937&     -- &    C1  \\ 
  1679 & 12.20 &  0.75 &\nodata &    5 &\nodata&\nodata&        -- &10.20 & 0.48 & 0.13 &        1 &       K2 &     1&     18-20 &     1&     -- &    --&     -- &    C1  \\ 
  1683 & 10.93 &  0.46 & 0.38 &    1 &10.44 & 0.48 &        3 &10.05 & 0.12 & 0.05 &        1 &    A1,A0 &   1,3&       $<$50 &     1&    464 &    --&     -- &    C3  \\ 
  1685 & 10.19 &  0.15 & 0.14 &    1 & 9.89 & 0.27 &        1 & 9.39 & 0.26 & 0.25 &        1 &       B9 &   1,2&  201,240/50 &   1,9&   466  &    --&     -- &     D  \\ 
  1691 & 11.23 &  0.70 & 0.22 &  4,5 &10.42 & 0.81 &        4 & 9.87 & 0.39 & 0.08 &        1 &       G4 &     6&        68 &     7&     -- &    --&     -- &     C  \\ 
  1698 &  8.87 &  0.06 & 0.01 &    1 &\nodata &\nodata&        -- & 8.76 & 0.01 & 0.01 &        1 &       A1 & 1,3,11&       $<$50 &     1&    472 & 36957&     -- &    C1  \\ 
  1708 &  7.35 & -0.08 &-0.59 &    1 &\nodata &\nodata&        -- & 7.44 &-0.03 &-0.01 &        1 & B3-B5,B3 & 1,3,11&       $<$50 &     1&   480  & 36958&     -- &    C2  \\ 
  1712 & 10.47 &  0.57 & 0.25 &    1 & 9.74 & 0.70 &        3 & 9.15 & 0.23 & 0.13 &        1 &       B9 &   1,3&        75 &     1&    479 &    --&     -- &    C3  \\ 
  1716 &  5.71 & -0.20 &-0.90 &    1 &\nodata &\nodata&        -- &\nodata &\nodata&\nodata&        -- &       B1 & 12,11,3&    35,$<$10 & 12,10&    482 & 36959&   1886 &    C4  \\ 
  1728 &  4.81 & -0.25 &-1.01 &    1 &\nodata &\nodata&        -- &\nodata &\nodata&\nodata&        -- &     B0.5 & 12,11,3&     25,30 & 12,10&    493 & 36960&   1887 &    C4  \\ 
  1736 & 11.11 &  1.27 & 0.74 &    1 & 9.43 & 1.63 &        1 & 8.26 & 0.54 & 0.26 &        2 &    F8,G2 &   1,2&    41,100 &   1,7&    497 &    --&     -- &    C3  \\ 
  1744 &  7.84 & -0.12 &-0.56 &    1 & 8.05 &-0.16 &        1 & 8.12 &-0.06 &-0.04 &        1 &    B5,B4 & 1,11,2&  75,135,145 & 1,9,10&   502  & 36981&     -- &    C3  \\ 
  1746 & 11.66 &  1.08 &\nodata &    1 &10.18 & 1.34 &        1 & 9.03 & 0.90 & 0.79 &        2 &    K0,K2 &   1,2&     36,39 &   1,3&   510  &    --&     -- &     D  \\ 
  1768 &  9.23 & -0.03 &-0.12 &    1 & 9.21 & 0.02 &        4 & 9.20 & 0.02 & 0.03 &        1 &  B9,B9.5 & 1,3,11&       180 &     1&    520 & 36983&     -- &    C4  \\ 
  1772 &  8.46 &  0.09 &-0.61 &    1 & 8.04 & 0.33 &        1 & 7.75 & 0.13 & 0.10 &        1 & B2.5,1.5p & 1,2,11&    80,135 &   1,9&   530  & 36982&     -- &     D  \\ 
  1785 & 12.47 &  1.13 &\nodata &    3 &11.16 & 1.53 &        1 & 9.77 & 0.88 & 0.65 &        2 &       K0 &     1&     18-20,22 & 1,14 &    535 &    --&     -- &     D  \\ 
  1789 & 10.53 &  0.53 & 0.01 &    1 & 9.92 & 0.61 &        4 & 9.51 & 0.24 & 0.04 &        1 &       F6 &   1,3&        81 &     1&    540 &    --&     -- &    C4  \\ 
  1792 &  8.88 & -0.05 &-0.26 &    2 & 8.92 &-0.04 &        -- & 8.87 & 0.07 & 0.01 &        1 & B8-A0,A0 & 10,11&        75 &     8&     -- & 37001&     -- &     C  \\ 
  1795 &  9.00 & -0.01 &-0.15 &    1 & 9.00 & 0.00 &        4 & 8.98 & 0.03 &-0.03 &        1 &       B9 &  1,11&      350: &     1&    529 & 36998&     -- &    C1  \\ 
  1798 &  9.47 &  0.36 &-0.41 &    1 &\nodata &\nodata&        -- & 8.11 & 0.36 & 0.18 &        1 &       B3 & 1,3,11&       50: &     1&    545 & 294264&     -- &    C2  \\ 
  1799 & 12.64 &  1.14 &-0.11 &    1 &10.94 & 1.85 &        1 & 9.52 & 0.63 & 0.26 &        2 &    G0-G2 &     1&  123,10,220 & 1,3,5&   541  &    --&     -- &    D1  \\ 
  1813 &  7.49 & -0.14 &-0.66 &    1 &\nodata &\nodata&        -- & 7.69 & 0.01 &-0.02 &        1 & B4,B3,B5 & 1,3,11&        80 &     1&    552 & 37000&     -- &    C4  \\ 
  1828 & 12.44 &  1.14 & 0.63 &    3 &11.05 & 1.43 &        2 & 9.79 & 0.76 & 0.53 &        1 &       K3 &     1&        29 &     1&    563 &    --&     -- &    D1  \\ 
  1849 &  8.51 & -0.09 &-0.41 &    1 &\nodata &\nodata&        -- & 8.70 &-0.03 &-0.03 &        1 &    B8,B7 & 1,3,11&       50: &     1&    581 & 36999&     -- &    C4  \\ 
  1865 &  6.74 &  0.03 &-0.87 &    1 & 6.31 & 0.42 &        1 & 6.09 & 0.28 & 0.16 &        1 & O9,B0,O7 & 1,2,11&  $<$50,135 (SB1) &   1,9&    587 & 37020&   1893 &     D  \\ 
  1881 &  9.81 &  0.21 &-0.10 &    1 &\nodata &\nodata&        -- & 9.25 & 0.19 & 0.04 &        1 &    A0-A2 &   1,3&       $<$50 &     1&    599 & 294262&     -- &    C2  \\ 
  1891 &  5.14 & -0.01 &-0.95 &    1 & 4.81 & 0.33 &        1 & 4.63 & 0.15 & 0.07 &        1 &       O7e & 1,2,11&  100:,140,98 & 1,9,13&    598 & 37022&   1895 &     D  \\ 
  1905 &  9.39 &  0.05 & 0.05 &    1 & 9.36 & 0.04 &        3 & 9.24 & 0.03 & 0.04 &        1 &       A0 & 1,3,11&   177,190 &   1,9&    608 & 37019&     -- &    C3  \\ 
  1923 & 12.70 &  0.50 &\nodata &    5 &11.25 & 0.64 &        1 &10.03 & 0.56 & 0.33 &        2 &       A5 &   1,2&   125,$<$12 &   1,3&    633 &    --&     -- &     D  \\ 
  1929 & 13.30 &  0.48 &\nodata &    5 &\nodata&\nodata&        -- &10.05 & 0.76 & 0.42 &        1 &       K3 &     1&     34,50 &   1,3&    635 &    --&     -- &    D1  \\ 
  1933 &  6.57 & -0.14 &-0.77 &    1 &\nodata &\nodata&        -- & 4.96 & 0.08 &-0.01 &        1 &  B1.5-B3p & 11,12,1,3&   125,150 &  1,10&    632 & 37017&   1890 &    C1  \\ 
  1950 & 15.00 & -1.86 &\nodata &    5 &\nodata&\nodata&        -- &10.65 & 0.41 & 0.13 &        1 &       G8 &     1&        27 &     1&    645 &    --&     -- &    C2  \\ 
  1953 &  9.89 &  0.72 & 0.17 &    1 & 9.05 & 0.83 &        1 & 8.35 & 0.55 & 0.53 &        1 &    G1,G0 &   1,2&        31 &     1&   653  &    --&     -- &    C3  \\ 
  1955 & 10.91 &  1.09 & 0.60 &    1 & 9.65 & 1.33 &        1 & 8.75 & 0.59 & 0.21 &        1 &    G8:G2 &   1,2&   109,130 &   1,7&    656 &    --&     -- &    C3  \\ 
  1956 &  9.62 &  0.29 &-0.39 &    2 & 8.90 & 0.72 &        1 & 8.13 & 0.48 & 0.26 &        2 &    B3,B4 &   1,2&   200,190 &   1,9&    655 &    --&     -- &     D  \\ 
  1971 & 13.57 &  1.22 & 0.90 &    6 &12.25 & 1.36 &        1 &11.45 & 0.66 & 0.12 &        2 &    lateK &     2&        18 &     3&    671 &    --&     -- &    C3  \\ 
  1972 & 12.90 &  1.30 &\nodata &    5 &11.81 & 1.34 &        1 &10.91 & 0.61 & 0.13 &        2 &       K3 &   1,2&     18-20,12 & 1,14 &    676 &    --&     -- &     D  \\ 
  1973 & 12.80 &  1.50 &\nodata &    9 &11.83 & 1.44 &        1 &10.53 & 0.79 & 0.53 &        2 &       G9 &     2&        45 &     2&    669 &    --&     -- &     D  \\ 
  1993 &  5.06 & -0.08 &-0.97 &    1 & 4.97 & 0.10 &        1 & 4.95 & 0.07 & 0.01 &        1 &    O9-B0e & 1,2,11&  150,145 (SB1) & 1,9/10/13&    682 & 37041&   1897 &     D  \\ 
  1996 & 11.01 &  0.64 & 0.15 &    1 &10.21 & 0.78 &        5 & 9.68 & 0.34 & 0.02 &        1 &    F8,G5 &   1,3&     63,51 &   1,7&   684  &    --&     -- &    C4  \\ 
  2001 & 13.10 &  0.70 &\nodata &    5 &11.12 & 1.22 &        1 &10.23 & 0.61 & 0.12 &        2 &    K0,K1 &   1,2&      7-20 &     1&    690 &    --&     -- &    D1  \\ 
  2006 & 12.88 &  1.00 & 0.72 &    6 &12.49 & 1.34 &        1 &11.31 & 0.71 & 0.33 &        4 &       K4 &     2&        28 &     3&    696 &    --&     -- &    C3  \\ 
  2020 & 12.02 &  0.97 & 0.43 &    1 &10.88 & 1.06 &        1 &10.10 & 0.49 & 0.16 &        1 &    K2:K0 &   1,2&     36,38,42 &   1,3,14&   698  &    --&     -- &    C3  \\ 
  2031 &  6.41 & -0.11 &-0.93 &    1 & 6.41 & 0.00 &        1 & 6.46 & 0.07 & 0.07 &        2 &       B1 & 1,2,11&  50:,30,10 & 1,9,10&    714 & 37042&     -- &     D  \\ 
  2033 & 11.73 &  0.91 & 0.16 &    3 &10.59 & 1.19 &        1 & 9.73 & 0.55 & 0.19 &        2 &    K1:G5 &   1,2&   56,64   & 1,14 &    713 &    --&     -- &     D  \\ 
  2035 &  9.80 &  0.20 & 0.13 &    1 & 9.47 & 0.32 &        4 & 9.48 & 0.02 &-0.02 &        1 &       A3 &     1&       106 &     1&    720 &    --&     -- &    C4  \\ 
  2036 &  9.76 &  0.41 & 0.08 &    1 & 9.24 & 0.52 &        4 & 8.87 & 0.25 & 0.28 &        1 &       F2 &   1,3&  57 (SB:15) &     1&    718 &    --&     -- &    C4  \\ 
  2037 &  2.76 & -0.25 &-1.07 &    1 & 3.03 &-0.26 &        -- &3.31   &\nodata &\nodata&        -- &    O9III &  3,11&       120 & 13   &    721 & 37043&   1899 &    C4  \\ 
  2047 & 14.00 &  0.90 &\nodata &    9 &12.37 & 1.75 &        1 &11.00 & 0.73 & 0.24 &        2 &       M0 &     2&        12 &     3&    732 &    --&     -- &     D  \\ 
  2048 & 13.90 &  1.10 &\nodata &    9 &12.13 & 1.82 &        1 &10.85 & 0.70 & 0.15 &        2 &       K8 &     2&        14 &     3&    733 &    --&     -- &    D1  \\ 
  2058 &  9.57 &  0.04 & 0.03 &    1 & 9.47 & 0.00 &        1 & 9.41 & 0.07 & 0.02 &        2 &       A0 &   1,2&   170,200 &   1,9&    734 &    --&     -- &     D  \\ 
  2065 &  9.08 & -0.01 &-0.20 &    1 &\nodata &\nodata&        -- & 9.14 &-0.02 &-0.03 &        1 &    B9,B8 &   1,3&       180 &     1&    736 & 37059&     -- &    C2  \\ 
  2069 & 12.30 &  1.02 &\nodata &    5 &10.84 & 1.25 &        1 & 9.95 & 0.64 & 0.06 &        2 &       K3 &     1&     38,44 &   1,3&    744 &    --&     -- &    D1  \\ 
  2074 &  6.83 &  0.26 &-0.69 &    1 & 6.31 & 0.53 &        1 & 5.84 & 0.19 & 0.09 &        1 &    B2,B1 & 1,2,11&  225,160 (SB1) & 1,9/10&   747  & 37061&     -- &    D1  \\ 
  2083 &  7.30 & -0.13 &-0.78 &    1 &\nodata &\nodata&        -- & 7.66 &-0.02 &-0.03 &        1 &    B4,B3p &  1,11&   50:,$<$10 &  1,10&   761  & 37058&     -- &    C2  \\ 
  2084 & 12.49 &  1.31 & 0.40 &    1 &10.90 & 1.65 &        1 & 9.44 & 0.85 & 0.43 &        2 &    K3,K4 &   1,2&     19-20 &     1&   757  &    --&     -- &    C3  \\ 
  2085 &  8.21 &  0.02 &-0.47 &    1 & 8.03 & 0.21 &        1 & 7.88 & 0.19 & 0.10 &        2 &    B4,B5 & 12,11,1,2&   50:,140 &   1,9&   760  & 37062&     -- &     D  \\ 
  2086 &  9.95 &  0.50 & 0.22 &    1 & 9.29 & 0.57 &        1 & 8.81 & 0.54 & 0.61 &        1 &       F5 &   1,2&  72 (SB:25) &     1&   767  &    --&     -- &    D1  \\ 
  2100 & 11.80 &  0.94 &\nodata &    4 &10.30 & 1.42 &        1 & 9.32 & 0.67 & 0.17 &        2 &    K0:G0 &   1,2&     72,45,66 &   1,3,14&    773 &    --&     -- &    D1  \\ 
  2102 &  9.36 &  0.04 &-0.01 &    1 & 9.37 & 0.03 &        3 & 9.31 & 0.01 &-0.05 &        1 &       A0 & 1,3,11&   250,200 &   1,9&    776 & 37060&     -- &    C3  \\ 
  2118 &  9.90 &  0.08 & 0.07 &    1 & 9.80 & 0.08 &        1 & 9.22 & 0.56 & 0.27 &        1 &    A1,A0 &   1,2&  $<$50:,400/50 &   1,9&   786  &    --&     -- &    C3  \\ 
  2167 & 11.39 &  1.02 & 0.61 &    3 &10.20 & 1.10 &        1 & 9.37 & 0.54 & 0.18 &        2 &       K0 &   1,2&        57 &     1&    831 &    --&     -- &    D1  \\ 
  2216 & 12.08 &  1.04 & 0.45 &    6 &10.89 & 1.19 &        1 &10.06 & 0.60 & 0.29 &        1 &      K2: &     1&        54 &     1&    864 &    --&     -- &    C3  \\ 
  2244 & 12.35 &  1.25 & 0.46 &    1 &11.24 & 1.38 &        3 &10.09 & 0.64 & 0.16 &        1 &    K3,K1 &   1,4&     51,65 &   1,5&    887 &    --&     -- &    C3  \\ 
  2247 & 10.00 &  0.41 & 0.41 &    1 & 9.60 & 0.70 &        1 & 8.32 & 1.04 & 1.00 &        1 &       A3 &   1,2&       140 &     1&   884  &    --&     -- &    D1  \\ 
  2252 & 11.61 &  0.88 & 0.36 &    1 &10.52 & 1.00 &      5,3 & 9.65 & 0.61 & 0.43 &        1 &    G8,G7 & 1,3,4&     33,43 &   1,7&   892  &    --&     -- &    C3  \\ 
  2257 & 12.21 &  0.87 & 0.31 &    1 &11.27 & 0.93 &        4 &10.63 & 0.46 & 0.08 &        1 &      K0: &     1&     47,51 &   1,3&    900 &    --&     -- &    C2  \\ 
  2271 &  7.10 & -0.07 &-0.51 &    1 & 7.11 &-0.01 &        1 & 7.20 & 0.02 & 0.06 &        1 &      B6pe & 1,2,11&  250:,260/180 &   1,9&    907 & 37115&     -- &    D1  \\ 
  2284 &  9.03 &  0.01 &-0.11 &    1 & 9.07 &-0.08 &        1 & 8.99 &-0.02 & 0.01 &        1 &  B9.5,B8 & 1,2,11&   180,245 &   1,9&    920 & 37114&     -- &    D1  \\ 
  2305 & 13.30 &  1.40 &\nodata &    5 &11.79 & 1.57 &        1 &10.71 & 0.65 & 0.11 &        2 &    K6-K7 &     2&       $<$12 &     3&    935 &    --&     -- &    D1  \\ 
  2333 & 11.70 &  0.88 &\nodata &    5 &\nodata&\nodata&        -- &10.10 & 0.40 & 0.09 &        1 &       K0 &     1&    43,$<$12 &   1,3&    953 &    --&     -- &    D1  \\ 
  2346 & 10.96 &  0.52 & 0.05 &    1 &10.34 & 0.58 &      5,4 & 9.91 & 0.24 &-0.02 &        1 &       G0 &     3&        88 &     7&    961 &    --&     -- &    C1  \\ 
  2358 & 11.00 &  0.61 & 0.07 &    1 &10.25 & 0.71 &        5 & 9.74 & 0.28 & 0.06 &        1 &       G0 &     6&     35,58 &   3,7&    973 &    --&     -- &    C1  \\ 
  2366 &  6.57 & -0.17 &-0.79 &    1 &\nodata &\nodata&        -- & 7.00 &-0.07 &-0.07 &        1 &    B2,B3 & 12,1,3,11&   175,300 &   1,9&    980 & 37150&   1906 &    D1  \\ 
  2368 & 13.56 &  1.54 & 1.50 &    1 &\nodata &\nodata&        - &\nodata &\nodata& 0.15 &        4 &    K4,K6 &     3&        24 &    11&    982 &    --&     -- &    C3  \\ 
  2370 & 10.60 &  0.56 &\nodata &    5 & 9.77 & 0.71 &        5 & 9.41 & 0.32 & 0.08 &        1 &       G2 &     6&       160 &     7&     -- &    --&     -- &     C  \\ 
  2387 &  9.22 & -0.01 &-0.10 &    1 & 9.24 &-0.02 &        1 & 9.28 & 0.02 & 0.00 &        2 &    A0,B9 & 1,2,11&   125,190 &   1,9&   992  & 37174&     -- &    D1  \\ 
  2404 & 11.30 &  0.70 & 0.18 &    1 &10.48 & 0.76 &      5,4 &\nodata &\nodata &\nodata&        - &    G2,G1 &   1,5&     29,22 &   1,7&   1004 &    --&     -- &    D1  \\ 
  2412 & 12.82 &  0.98 & 0.67 &    8 &\nodata &\nodata&        -- &\nodata &\nodata& 0.57 &        4 &       K3 &     9&        47 &    11&     -- &    --&     -- &     C  \\ 
  2441 & 10.76 &  0.70 & 0.15 &    1 & 9.92 & 0.84 &      5,4 & 9.12 & 0.52 & 0.41 &        1 &       G2 &     3&        18 &     7&   1030 &    --&     -- &    C1  \\ 
  2486 & 11.37 &  0.61 & 0.04 &    1 &10.69 & 0.69 &        5 &10.22 & 0.27 & 0.07 &        1 &    G5,G7 &   3,6&       120 &     7&   1060 &    --&     -- &    C3  \\ 
  2494 & 10.80 &  0.90 & 0.43 &    1 & 9.75 & 0.99 &        5 & 9.03 & 0.46 & 0.12 &        1 & F7,G0,K0 & 6,3,4&        26 &     7&  1069  &    --&     -- &     C  \\ 
  2602 &  6.20 & -0.04 &-0.73 &    2 &\nodata &\nodata&        -- &\nodata &\nodata&\nodata&        -- &  B1.5,B2 & 12,11,7&     20,10 &  4,10&   1129 & 37356&   1923 &     C  \\ 
  2660 & 11.27 &  0.54 & 0.04 &    7 &10.65 & 0.62 &        5 &10.27 & 0.26 &-0.02 &        1 &       F6 &     6&       175 &     7&     -- &    --&     -- &     C  \\ 
  2711 &  5.96 & -0.24 &-0.92 &    2 &\nodata &\nodata&        -- & 6.50 &-0.09 &-0.08 &        1 &  B1.5,IV & 12,7,11&    60,105 &  4,10&     -- & 37481&   1933 &     C  \\ 
  2758 &  7.61 & -0.13 &-0.55 &    2 & 7.77 &-0.16 &        -- &\nodata &\nodata &\nodata&        -- &       B3 & 10,11&       130 &     8&     -- & 37526&     -- &     C  \\ 
  2921 &  8.01 & -0.10 &-0.47 &    2 & 8.18 &-0.17 &        -- &\nodata &\nodata &\nodata&        -- &    B5-B6 & 10,11&       125 &     8&     -- & 37700&     -- &     C  \\ 
\enddata

\tablenotetext{a}{UBV references:
(1) Walker, 1969;
(2) Warren and Hesser, 1977;
(3) Penston, 1973;
(4) McNamara, 1976;
(5) McNamara et al., 1989;
(6) Penston, Hunter, \& O'Neill, 1973;
(7) Rydgren \& Vrba, 1984;
(8) Mundt \& Bastien, 1980;
(9) Parenago, 1954.
}

\tablenotetext{b}{VI references:
(1) Hillenbrand, 1997;
(2) Penston, 1973 transformed from johnson system to cousins system;
(3) Penston, Hunter, \& O'Neill, 1973 transformed to cousins system;
(4) McNamara, 1976 transformed to cousins system;
(5) Rydgren \& Vrba, 1984.
}

\tablenotetext{c}{JHK references:
(1) this paper, 1991 nov/dec OTTO observations; 
(2) this paper, 1992 SQIID observations; 
(3) this paper, 1993 NICMASS observations;
(4) literature.
}

\tablenotetext{d}{Spectral Type references:
(1) this paper; 
(2) Hillenbrand, 1997;
(3) Walker, 1969;
(4) Walker, 1983;
(5) Duncan, 1993;
(6) Smith, Beckers, \& Barden, 1983;
(7) Wolff, Edwards, \& Preston, 1982; 
(8) Penston, Hunter, \& O'Neill, 1973;
(9) Cohen \& Kuhi, 1979;
(10) reference in Warren and Hesser, 1977; 
(11) Brown et al, 1994;
(12) Wolff, 1990.
}

\tablenotetext{e}{The notation (SB:$nn$) in this column
indicates a candidate spectroscopic binary with $nn$
corresponding to the amplitude of the radial velocity
variation among our different observations of the star.
(SB1) indicates an SB1 candidate identified in previous literature;
see vsini references column.
}

\tablenotetext{f}{vsini references:
(1) this paper;
(2) Hartmann, 1996, private communication;
(3) Duncan, 1993;
(4) Wolff, Edwards, \& Preston, 1982; 
(5) Walker, 1990;
(6) Walker, 1983;
(7) Smith, Beckers, \& Barden, 1983;
(8) McNamara, 1963 (old slettebak system);
(9) Abt, Muncaster, \& Thompson, 1970;
(10) McNamara \& Larsson, 1962 (old slettebak system);
(11) Hartmann et al, 1986;
(12) Abt, \& Hunter, 1962;
(13) Conti \& Ebbets, 1977;
(14) Rhode, Herbst, \& Mathieu, 2000.
}

\end{deluxetable}
\clearpage

%
%
%


\tablenum{2}
\pagestyle{empty}

\begin{deluxetable}{c r c r r r r r c r r r}
\tabletypesize{\scriptsize}
\tablecaption{Derived Parameters} 
\tablehead{
\colhead{ Par. }&
\colhead{ vsini  }&
\colhead{ vsini \tablenotemark{a}}&
\colhead{ A$_V$ }&
\colhead{ log T$_{eff}$ }&
\colhead{ log L}&
\colhead{ R }&
\colhead{ log I }&
\colhead{ conv./rad. \tablenotemark{a} }&
\colhead{ log A}&
\colhead{ M }&
\colhead{ $\Delta$(H-K)}  \\

      \colhead{ }&
      \colhead{ adopted}&
      \colhead{ ref}&
      \colhead{ mag } &
      \colhead{ K } &
      \colhead{ L$_\odot$}&
      \colhead{ R$_\odot$}&
      \colhead{ g cm$^2$}&
      \colhead{ }&
      \colhead{ yr } &
      \colhead{ M$_\odot$} &
      \colhead{ mag}  
}
\startdata

    82 &   200 &     5 &  0.10 & 4.037 & 2.15 & 3.40 & 55.43 &  R & 6.15 & 3.74 &\nodata \\ 
   378 &    15 &     5 &  0.08 & 4.025 & 1.84 & 2.51 & 54.98 &  R & 6.46 & 2.90 &-0.03 \\ 
   597 &    25 &     5 &  0.42 & 4.025 & 2.16 & 3.65 & 55.46 &  R & 6.13 & 3.75 &\nodata \\ 
   679 &    15 &     3 &  0.17 & 4.276 & 3.43 & 4.94 & 55.77 &  R &\nodata & 8.01 &-0.04 \\ 
   854 &   190 &     5 &  0.20 & 4.130 & 2.56 & 3.54 & 55.49 &  R & 6.03 & 4.45 &-0.03 \\ 
   908 &   135 &     5 &  0.10 & 4.037 & 1.81 & 2.31 & 54.91 &  R & 6.51 & 2.83 &-0.04 \\ 
  1044 &   $<$50 &       &  0.79 & 4.294 & 3.13 & 3.21 & 55.55 &  R &\nodata & 6.47 &-0.01 \\ 
  1049 &    20 &       &  1.20 & 3.623 & 1.14 & 7.18 & 55.78 &  C & 4.52 & 0.71 & 0.70 \\ 
  1076 &    22 &     7 &  0.17 & 3.708 & 0.21 & 1.67 & 54.71 &  R & 6.84 & 1.39 & 0.36 \\ 
  1097 &   150 &     5 &  0.14 & 4.053 & 1.95 & 2.50 & 55.05 &  R & 6.39 & 3.14 & 0.00 \\ 
  1126 &    80 &       &  0.30 & 4.025 & 1.79 & 2.38 & 54.92 &  R & 6.52 & 2.80 &-0.04 \\ 
  1179 &   114 &       &  0.14 & 3.790 & 0.55 & 1.68 & 54.29 &  R & 7.11 & 1.41 &-0.02 \\ 
  1270 &    16 &     7 &  0.34 & 3.708 & 0.48 & 2.26 & 55.19 &C/R & 6.47 & 1.71 & 0.17 \\ 
  1319 &   187 &       &  1.60 & 3.761 & 0.72 & 2.36 & 54.84 &  R & 6.80 & 1.72 &-0.00 \\ 
  1322 &    70 &       &  0.49 & 3.736 & 0.66 & 2.45 & 55.14 &  R & 6.56 & 1.91 & 0.01 \\ 
  1326 &    20 &       &  1.08 & 3.643 & 1.24 & 7.34 & 55.97 &  C & 4.62 & 0.96 & 0.00 \\ 
  1345 &    20 &       &  0.46 & 3.643 & 0.73 & 4.07 & 55.40 &  C & 5.57 & 0.84 & 0.04 \\ 
  1360 &    15 &     2 &  0.77 & 3.736 &-0.07 & 1.05 & 53.98 &  R & 7.46 & 1.00 &\nodata \\ 
  1374 &    77 &       &  0.32 & 3.798 & 1.25 & 3.62 & 55.07 &  R & 6.59 & 2.22 &-0.01 \\ 
  1391 &    20 &       &  0.17 & 3.788 & 0.92 & 2.59 & 54.71 &  R & 6.86 & 1.74 & 0.01 \\ 
  1393 &    20 &       &  0.42 & 3.748 & 0.58 & 2.11 & 54.82 &  R & 6.81 & 1.65 &-0.00 \\ 
  1394 &    60 &       &  0.28 & 3.798 & 1.24 & 3.60 & 55.07 &  R & 6.59 & 2.21 & 0.25 \\ 
  1404 &    34 &     2 &  0.67 & 3.753 & 0.93 & 3.10 & 55.29 &  R & 6.49 & 2.19 & 0.37 \\ 
  1408 &    21 &       &  0.17 & 3.623 & 0.36 & 1.65 & 55.01 &  C & 5.78 & 0.68 & 0.02 \\ 
  1409 &    29 &       &  0.91 & 3.771 & 0.83 & 2.55 & 54.84 &  R & 6.80 & 1.78 & 0.47 \\ 
  1414 &    29 &       &  0.20 & 3.766 & 0.60 & 2.00 & 54.55 &  R & 6.98 & 1.48 &-0.03 \\ 
  1425 &    26 &       &  0.75 & 3.742 & 0.64 & 2.32 & 55.03 &  R & 6.66 & 1.82 & 0.04 \\ 
  1440 &    20 &       &  0.17 & 3.661 & 0.27 & 2.21 & 54.99 &  C & 6.23 & 1.12 &-0.04 \\ 
  1445 &   280 &       &  0.13 & 4.111 & 2.31 & 2.90 & 55.33 &  R & 6.15 & 3.94 &-0.02 \\ 
  1455 &    21 &       &  0.26 & 3.771 & 1.00 & 3.10 & 55.10 &  R & 6.61 & 2.07 & 0.05 \\ 
  1484 &    49 &       &  0.84 & 3.708 & 0.79 & 3.22 & 55.56 &  C & 6.07 & 1.96 &-0.02 \\ 
  1491 &   225 &       &  0.08 & 4.053 & 2.40 & 4.21 & 55.55 &  R & 6.05 & 4.10 & 0.01 \\ 
  1505 &    20 &       &  0.01 & 3.695 & 0.10 & 1.56 & 54.67 &  R & 6.86 & 1.31 & 0.01 \\ 
  1507 &   136 &       &  0.45 & 3.907 & 1.16 & 1.99 & 54.40 &  R & 6.93 & 1.82 &-0.01 \\ 
  1510 &    38 &       &  1.19 & 3.695 & 0.87 & 3.77 & 55.66 &  C & 5.83 & 1.72 & 0.03 \\ 
  1511 &   112 &       &  0.40 & 3.960 & 1.73 & 2.98 & 55.06 &  R & 6.48 & 2.75 &-0.04 \\ 
  1518 &    38 &     2 &  0.17 & 3.695 &-0.10 & 1.24 & 55.29 &  R & 7.18 & 1.09 & 0.15 \\ 
  1539 &   250 &       &  2.47 & 4.053 & 2.02 & 2.72 & 55.17 &  R & 6.30 & 3.37 &-0.05 \\ 
  1540 &    20 &       &  0.76 & 3.661 & 1.20 & 6.46 & 55.94 &  C & 5.28 & 1.23 & 0.09 \\ 
  1541 &    20 &       &  0.38 & 3.695 & 0.48 & 2.40 & 55.21 &  C & 6.31 & 1.58 & 0.06 \\ 
  1552 &    15 &     2 &  0.17 & 3.602 & 0.21 & 2.71 & 54.84 &  C & 5.84 & 0.55 & 0.56 \\ 
  1553 &    28 &       &  1.53 & 3.695 & 1.01 & 4.41 & 55.81 &  C & 5.68 & 1.78 & 0.29 \\ 
  1554 &    30 &       &  0.28 & 3.713 & 0.37 & 1.95 & 54.94 &  R & 6.66 & 1.58 &-0.00 \\ 
  1562 &   200 &       &  0.46 & 4.012 & 1.63 & 2.10 & 54.70 &  R & 6.67 & 2.46 &-0.05 \\ 
  1581 &    20 &       &  0.93 & 3.806 & 0.49 & 1.46 & 54.18 &  R & 7.21 & 1.36 &-0.00 \\ 
  1587 &    20 &       &  1.34 & 3.695 & 0.85 & 3.69 & 55.63 &  C & 5.84 & 1.71 & 0.04 \\ 
  1605 &   127 &       &  0.29 & 3.933 & 2.16 & 5.54 & 55.70 &  R & 6.00 & 3.78 & 0.48 \\ 
  1608 &    26 &       &  0.17 & 3.652 & 0.20 & 2.14 & 54.93 &  C & 6.22 & 1.02 &-0.01 \\ 
  1623 &   260 &       &  1.58 & 3.940 & 1.70 & 3.15 & 55.08 &  R & 6.48 & 2.72 & 0.38 \\ 
  1626 &    94 &       &  1.76 & 3.917 & 1.30 & 2.23 & 54.55 &  R & 6.83 & 1.98 &-0.05 \\ 
  1634 &   150 &       &  0.04 & 4.086 & 2.11 & 2.58 & 55.18 &  R & 6.27 & 3.51 &-0.03 \\ 
  1643 &    20 &       &  0.17 & 3.719 & 0.06 & 1.33 & 54.35 &  R & 7.16 & 1.17 & 0.10 \\ 
  1646 &    20 &       &  0.10 & 3.798 & 1.22 & 3.51 & 55.03 &  R & 6.62 & 2.16 &-0.07 \\ 
  1654 &   120 &       &  0.51 & 4.053 & 1.99 & 2.65 & 55.14 &  R & 6.33 & 3.29 & 0.01 \\ 
  1657 &    20 &       &  0.17 & 3.778 & 0.57 & 1.82 & 54.42 &  R & 7.06 & 1.43 &-0.02 \\ 
  1659 &    20 &       &  0.84 & 3.679 & 1.08 & 5.16 & 55.84 &  C & 5.47 & 1.46 & 0.16 \\ 
  1660 &   275 &       &  0.20 & 4.025 & 1.89 & 2.67 & 55.09 &  R & 6.40 & 3.07 &-0.00 \\ 
  1664 &  180: &       &  0.21 & 4.210 & 2.76 & 3.10 & 55.45 &  R & 5.98 & 4.93 & 0.05 \\ 
  1671 &   175 &       &  0.31 & 3.897 & 1.35 & 2.58 & 54.66 &  R & 6.76 & 2.04 &-0.04 \\ 
  1679 &    20 &       &  0.17 & 3.695 & 0.40 & 2.19 & 55.12 &  C & 6.42 & 1.55 & 0.04 \\ 
  1683 &   $<$50 &       &  1.51 & 3.985 & 1.41 & 1.84 & 55.09 &  R &\nodata & 2.58 &-0.04 \\ 
  1685 &   201 &       &  0.73 & 4.025 & 1.62 & 1.97 & 54.67 &  R & 6.69 & 2.47 & 0.22 \\ 
  1691 &    68 &     4 &  0.26 & 3.755 & 0.73 & 2.44 & 54.97 &  R & 6.71 & 1.82 & 0.01 \\ 
  1698 &   $<$50 &       &  0.24 & 3.980 & 1.72 & 2.69 & 54.96 &  R & 6.53 & 2.68 &-0.01 \\ 
  1708 &   $<$50 &       &  0.47 & 4.272 & 3.09 & 3.40 & 55.52 &  R &\nodata & 6.32 & 0.00 \\ 
  1712 &    75 &       &  2.03 & 4.025 & 1.89 & 2.67 & 55.09 &  R & 6.40 & 3.07 & 0.01 \\ 
  1716 &   $<$10 &     6 &  0.24 & 4.377 & 3.88 & 5.24 & 56.14 &  R &\nodata &11.30 &\nodata \\ 
  1728 &    30 &     6 &  0.15 & 4.439 & 4.35 & 6.73 & 56.56 &  R &\nodata &16.63 &\nodata \\ 
  1736 &    41 &       &  2.10 & 3.761 & 1.51 & 5.83 & 56.03 &  R & 5.90 & 3.43 & 0.08 \\ 
  1744 &    75 &       &  0.21 & 4.195 & 2.63 & 2.86 & 55.37 &  R & 6.05 & 4.64 &-0.02 \\ 
  1746 &    36 &       &  0.76 & 3.702 & 0.98 & 4.14 & 55.79 &  C & 5.80 & 1.97 & 0.66 \\ 
  1768 &   180 &       &  0.17 & 4.025 & 1.64 & 2.01 & 54.69 &  R & 6.67 & 2.50 & 0.03 \\ 
  1772 &    80 &       &  1.03 & 4.294 & 3.06 & 2.98 & 55.50 &  R &\nodata & 6.19 & 0.08 \\ 
  1785 &    20 &       &  1.14 & 3.719 & 0.78 & 3.04 & 55.50 &  C & 6.22 & 2.11 & 0.51 \\ 
  1789 &    81 &       &  0.22 & 3.798 & 0.98 & 2.65 & 54.70 &  R & 6.84 & 1.79 &-0.01 \\ 
  1792 &    75 &     5 &  0.11 & 4.025 & 1.76 & 2.29 & 54.87 &  R & 6.55 & 2.73 & 0.01 \\ 
  1795 &  350: &       &  0.23 & 4.025 & 1.76 & 2.29 & 54.87 &  R & 6.55 & 2.73 &-0.03 \\ 
  1798 &   50: &       &  1.84 & 4.272 & 2.79 & 2.40 & 55.31 &  R &\nodata & 5.17 & 0.11 \\ 
  1799 &   123 &       &  1.75 & 3.766 & 0.90 & 2.82 & 55.02 &  R & 6.66 & 1.95 & 0.10 \\ 
  1813 &    80 &       &  0.23 & 4.240 & 2.87 & 3.07 & 55.37 &  R &\nodata & 5.46 & 0.00 \\ 
  1828 &    29 &       &  0.62 & 3.679 & 0.66 & 3.20 & 55.40 &  C & 5.89 & 1.35 & 0.40 \\ 
  1849 &   50: &       &  0.03 & 4.070 & 2.00 & 2.47 & 55.08 &  R & 6.36 & 3.26 &-0.02 \\ 
  1865 &   $<$50 &       &  1.04 & 4.471 & 4.15 & 4.64 & 56.38 &  R &\nodata &14.08 & 0.14 \\ 
  1881 &   $<$50 &       &  0.77 & 3.993 & 1.58 & 2.16 & 54.67 &  R &\nodata & 2.36 &-0.01 \\ 
  1891 &  100: &       &  0.97 & 4.603 & 5.09 & 7.44 & 57.34 &  R &\nodata &33.36 & 0.06 \\ 
  1905 &   177 &       &  0.28 & 3.993 & 1.55 & 2.09 & 54.73 &  R &\nodata & 2.43 & 0.03 \\ 
  1923 &   125 &       &  1.20 & 3.917 & 1.00 & 1.04 & 54.27 &  R & 7.67 & 1.74 & 0.24 \\ 
  1929 &    34 &       &  0.17 & 3.679 & 0.14 & 1.75 & 54.87 &C/R & 6.65 & 1.32 & 0.32 \\ 
  1933 &   125 &       &  0.34 & 4.307 & 3.42 & 4.26 & 55.77 &  R &\nodata & 7.98 & 0.01 \\ 
  1950 &    27 &       &  0.17 & 3.736 & 1.31 & 0.55 & 56.04 &  R & 5.94 & 3.12 & 0.06 \\ 
  1953 &    31 &       &  0.51 & 3.771 & 1.36 & 4.66 & 55.57 &  R & 6.29 & 2.74 & 0.45 \\ 
  1955 &   109 &       &  1.54 & 3.761 & 1.37 & 4.94 & 55.68 &  R & 6.22 & 2.83 & 0.06 \\ 
  1956 &   200 &       &  1.59 & 4.253 & 2.73 & 2.46 & 55.34 &  R &\nodata & 4.99 & 0.20 \\ 
  1971 &    18 &     2 &  0.02 & 3.623 &-0.01 & 1.91 & 54.72 &  C & 6.26 & 0.80 & 0.01 \\ 
  1972 &    20 &       &  1.12 & 3.679 & 0.68 & 3.25 & 55.40 &  C & 5.87 & 1.35 &-0.03 \\ 
  1973 &    45 &     1 &  2.32 & 3.722 & 1.12 & 4.42 & 55.92 &  C & 5.92 & 2.65 & 0.31 \\ 
  1993 &   150 &       &  0.74 & 4.543 & 4.88 & 7.69 & 57.11 &  R &\nodata &27.07 & 0.01 \\ 
  1996 &    63 &       &  0.20 & 3.766 & 0.79 & 2.48 & 54.86 &  R & 6.79 & 1.77 &-0.04 \\ 
  2001 &    20 &       &  0.17 & 3.708 & 0.16 & 1.56 & 54.59 &  R & 6.93 & 1.31 & 0.03 \\ 
  2006 &    28 &     2 &  0.17 & 3.661 & 0.21 & 2.06 & 54.94 &  C & 6.33 & 1.15 & 0.22 \\ 
  2020 &    36 &       &  0.64 & 3.719 & 0.62 & 2.52 & 55.25 &  R & 6.42 & 1.89 & 0.05 \\ 
  2031 &   50: &       &  0.53 & 4.383 & 3.88 & 5.05 & 56.14 &  R &\nodata &11.23 & 0.09 \\ 
  2033 &    56 &       &  0.78 & 3.744 & 0.90 & 3.10 & 55.37 &  R & 6.42 & 2.22 & 0.08 \\ 
  2035 &   106 &       &  0.47 & 3.945 & 1.39 & 2.16 & 54.54 &  R & 6.82 & 2.06 &-0.06 \\ 
  2036 &    57 &       &  0.14 & 3.826 & 1.26 & 3.22 & 54.88 &  R & 6.68 & 2.09 & 0.24 \\ 
  2037 &   120 &     8 &  0.22 & 4.555 & 5.48 &14.46 & 57.80 &  R &\nodata &49.83 &\nodata \\ 
  2047 &    12 &     2 &  0.17 & 3.580 & 0.19 & 2.92 & 54.78 &  C & 5.74 & 0.41 & 0.03 \\ 
  2048 &    14 &     2 &  0.17 & 3.591 & 0.18 & 2.76 & 54.80 &  C & 5.82 & 0.48 &-0.04 \\ 
  2058 &   170 &       &  0.24 & 3.993 & 1.61 & 2.24 & 54.69 &  R & 6.68 & 2.38 & 0.01 \\ 
  2065 &   180 &       &  0.19 & 4.037 & 1.74 & 2.13 & 54.81 &  R & 6.58 & 2.70 &-0.03 \\ 
  2069 &    38 &       &  0.25 & 3.679 & 0.57 & 2.88 & 55.30 &  C & 5.99 & 1.34 &-0.04 \\ 
  2074 &   225 &       &  1.68 & 4.383 & 4.17 & 7.06 & 56.39 &  R &\nodata &14.23 & 0.03 \\ 
  2083 &   50: &       &  0.19 & 4.200 & 2.85 & 3.59 & 55.36 &  R &\nodata & 5.38 &-0.01 \\ 
  2084 &    20 &       &  0.88 & 3.661 & 0.65 & 3.43 & 55.36 &  C & 5.73 & 1.06 & 0.28 \\ 
  2085 &   50: &       &  0.68 & 4.210 & 2.84 & 3.40 & 55.35 &  R &\nodata & 5.36 & 0.09 \\ 
  2086 &    72 &       &  0.21 & 3.806 & 1.35 & 3.93 & 55.15 &  R & 6.53 & 2.35 & 0.56 \\ 
  2100 &    72 &       &  1.19 & 3.771 & 1.01 & 3.12 & 55.11 &  R & 6.60 & 2.09 & 0.05 \\ 
  2102 &   250 &       &  0.24 & 3.993 & 1.55 & 2.09 & 54.73 &  R &\nodata & 2.43 &-0.06 \\ 
  2118 &  $<$50: &       &  0.37 & 3.993 & 1.38 & 1.72 & 54.99 &  R &\nodata & 2.77 & 0.25 \\ 
  2167 &    57 &       &  0.80 & 3.719 & 1.07 & 4.27 & 55.90 &  C & 5.92 & 2.53 & 0.06 \\ 
  2216 &    54 &       &  0.54 & 3.695 & 0.59 & 2.74 & 55.34 &  C & 6.15 & 1.62 & 0.17 \\ 
  2244 &    51 &       &  1.19 & 3.695 & 0.75 & 3.27 & 55.60 &  C & 5.96 & 1.67 & 0.00 \\ 
  2247 &   140 &       &  1.08 & 3.940 & 1.69 & 3.15 & 55.08 &  R & 6.48 & 2.71 & 0.92 \\ 
  2252 &    33 &       &  0.58 & 3.736 & 0.73 & 2.66 & 55.26 &  R & 6.48 & 2.03 & 0.33 \\ 
  2257 &    47 &       &  0.33 & 3.719 & 0.42 & 2.00 & 54.96 &  R & 6.66 & 1.63 &-0.01 \\ 
  2271 &  250: &       &  0.19 & 4.111 & 2.90 & 5.71 & 55.39 &  R &\nodata & 5.55 & 0.07 \\ 
  2284 &   180 &       &  0.30 & 4.053 & 1.99 & 2.64 & 55.14 &  R & 6.33 & 3.29 & 0.00 \\ 
  2305 &   $<$12 &     2 &  0.40 & 3.612 & 0.43 & 3.34 & 55.05 &  C & 5.67 & 0.58 &-0.04 \\ 
  2333 &    43 &       &  0.36 & 3.719 & 0.78 & 3.03 & 55.50 &  C & 6.23 & 2.10 &-0.01 \\ 
  2346 &    88 &     4 &  0.17 & 3.771 & 0.79 & 2.43 & 54.77 &  R & 6.84 & 1.71 &-0.08 \\ 
  2358 &    35 &     2 &  0.17 & 3.771 & 0.78 & 2.39 & 54.75 &  R & 6.85 & 1.69 & 0.00 \\ 
  2366 &   175 &       &  0.25 & 4.309 & 3.54 & 4.80 & 55.86 &  R &\nodata & 8.66 &-0.04 \\ 
  2368 &    24 &     7 &  1.32 & 3.643 & 0.45 & 2.96 & 55.12 &  C & 5.82 & 0.85 &-0.04 \\ 
  2370 &   160 &     4 &  0.17 & 3.761 & 0.94 & 3.03 & 55.19 &  R & 6.55 & 2.12 & 0.02 \\ 
  2387 &   125 &       &  0.17 & 4.010 & 1.75 & 2.44 & 54.90 &  R & 6.53 & 2.73 &-0.00 \\ 
  2404 &    29 &       &  0.33 & 3.761 & 0.87 & 2.79 & 55.08 &  R & 6.63 & 1.98 &\nodata \\ 
  2412 &    47 &     7 &  0.12 & 3.679 & 0.17 & 1.81 & 54.93 &C/R & 6.60 & 1.33 & 0.47 \\ 
  2441 &    18 &     4 &  0.33 & 3.761 & 0.94 & 3.03 & 55.20 &  R & 6.55 & 2.12 & 0.34 \\ 
  2486 &   120 &     4 &  0.17 & 3.745 & 0.65 & 2.33 & 55.01 &  R & 6.68 & 1.82 & 0.00 \\ 
  2494 &    26 &     4 &  0.85 & 3.753 & 1.15 & 3.96 & 55.55 &  R & 6.31 & 2.55 & 0.01 \\ 
  2602 &    20 &     3 &  0.68 & 4.332 & 3.76 & 5.62 & 56.04 &  R &\nodata &10.29 &\nodata \\ 
  2660 &   175 &     4 &  0.22 & 3.795 & 0.68 & 1.91 & 54.39 &  R & 7.05 & 1.48 &-0.07 \\ 
  2711 &    60 &     3 &  0.09 & 4.352 & 3.67 & 4.58 & 55.96 &  R &\nodata & 9.55 &-0.04 \\ 
  2758 &   130 &     5 &  0.32 & 4.272 & 2.92 & 2.81 & 55.41 &  R &\nodata & 5.65 &\nodata \\ 
  2921 &   125 &     5 &  0.13 & 4.125 & 2.39 & 2.99 & 55.36 &  R & 6.12 & 4.10 &\nodata \\

\enddata

\tablenotetext{a}{vsini references: this paper unless otherwise indicated as
(1) Hartmann, 1986;
(2) Duncan, 1993;
(3) Wolff, Edwards, \& Preston 1982;
(4) Smith, Beckers, \& Barden 1983;
(5) McNamara, 1963;
(6) McNamara \& Larsson 1962;
(7) Hartmann et al. 1996, private communication;
(8) Conti \& Ebbets 1977;
}

\tablenotetext{b}{C/R indicate whether star is on convective vs radiative track in the HR diagram.}

\end{deluxetable}
\clearpage


\clearpage
\begin{figure}
\plotone{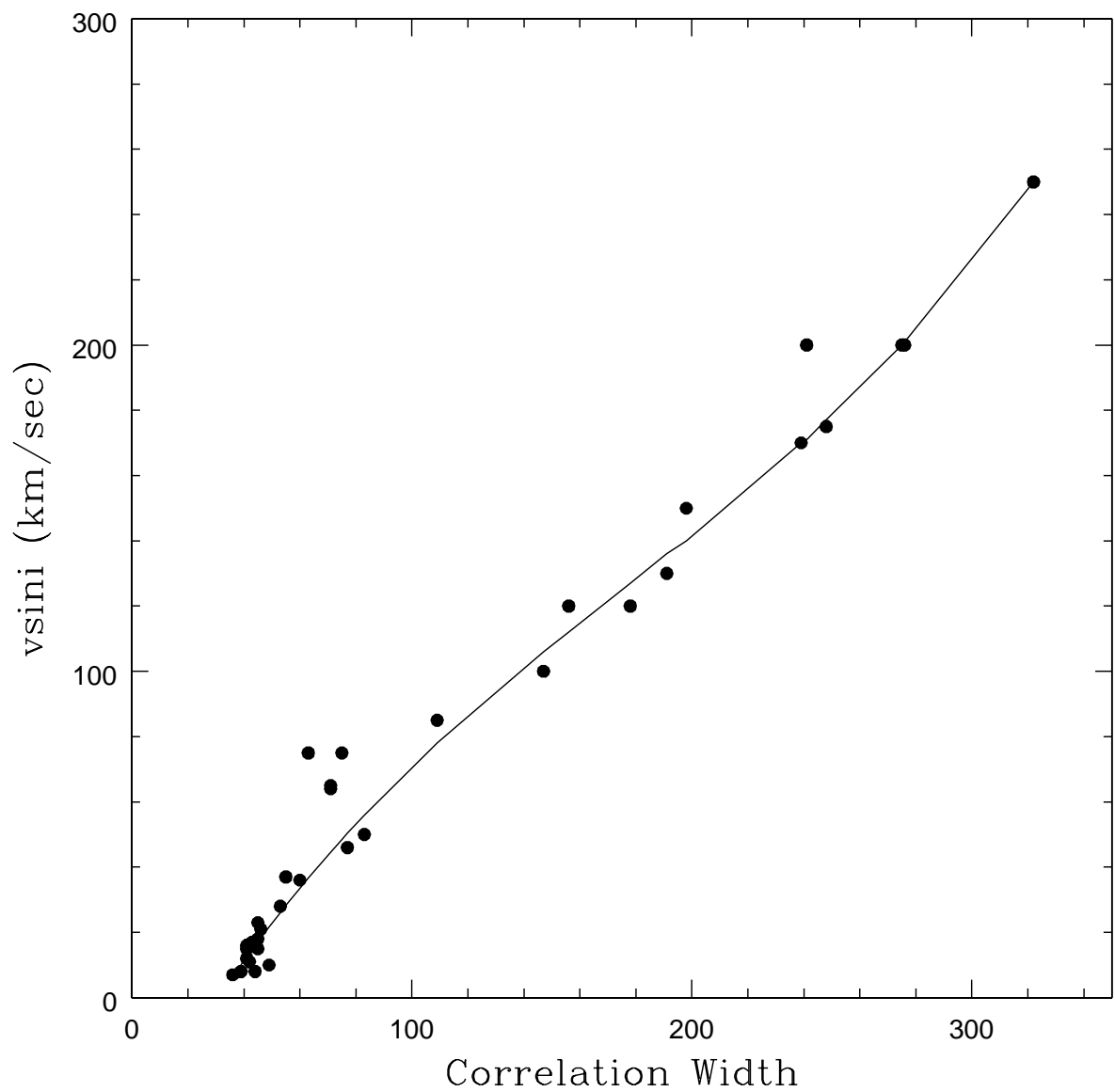}
\figcaption{
Full-width half-maximum of the cross-correlation peak 
vs. observed values of \textit{vsini} for standard stars used to establish 
functional relationship. The solid line shows the polynomial fit to the 
data.
}
\end{figure}

\clearpage
\begin{figure}
\plotone{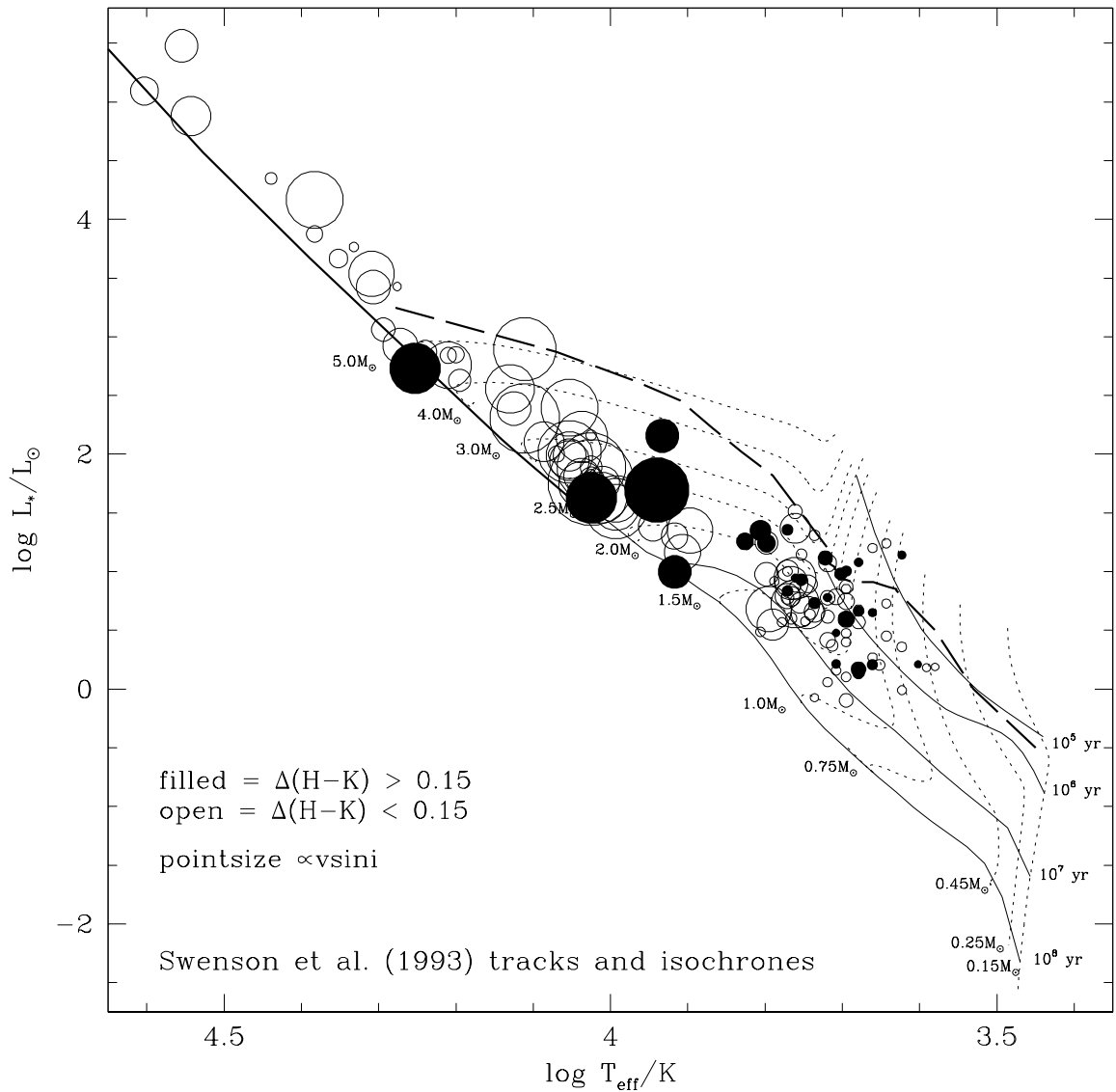}
\figcaption{
The HR diagram for stars in the current sample. Circle size 
is proportional to the observed rotation (\textit{vsini}). Filled 
circles represent stars with infrared excesses, which are diagnostic 
of inner accretion disks. The light dashed lines are evolutionary tracks from 
SFRI. The heavy line with long dashes shows the location of the 
PS birthline for an accretion rate of 10$^{-5}$ M$_{\odot}$ per year (Palla 
\& Stahler 1993). The solid lines are the isochrones of SFRI from 
10$^5$-10$^8$ yr (uncorrected for the birthline).
}
\end{figure}

\clearpage
\begin{figure}
\plottwo{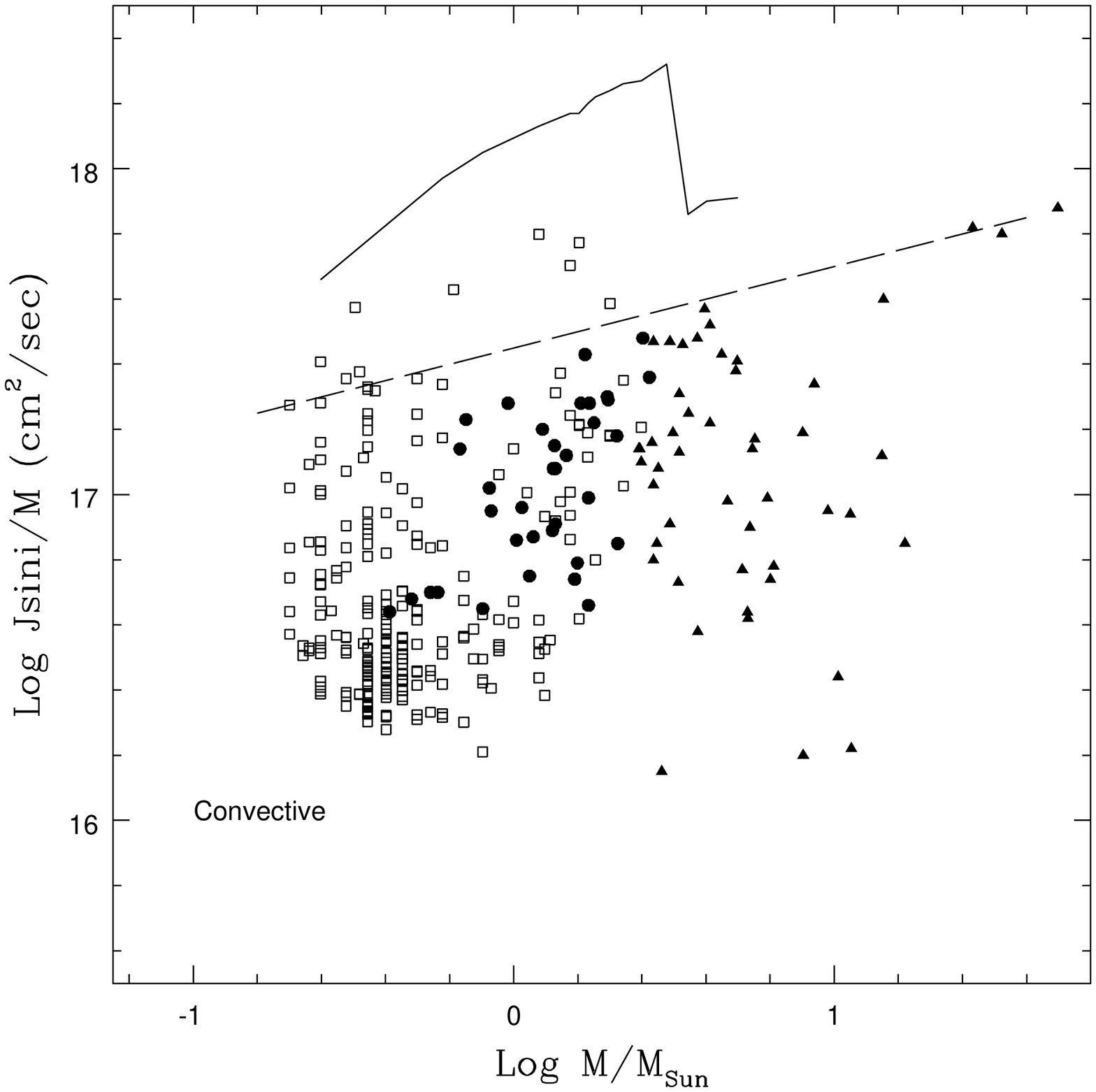}{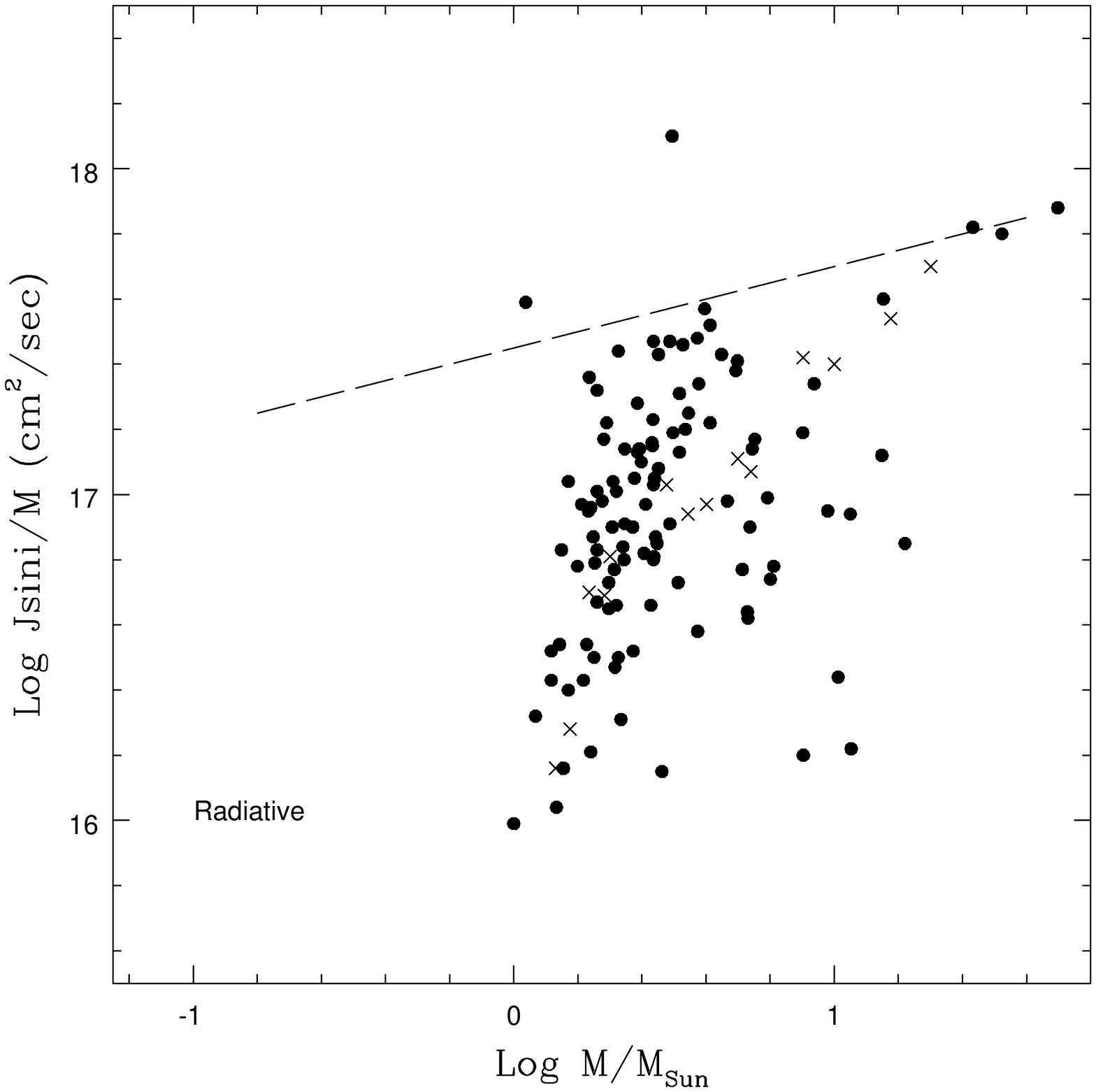}
\plotone{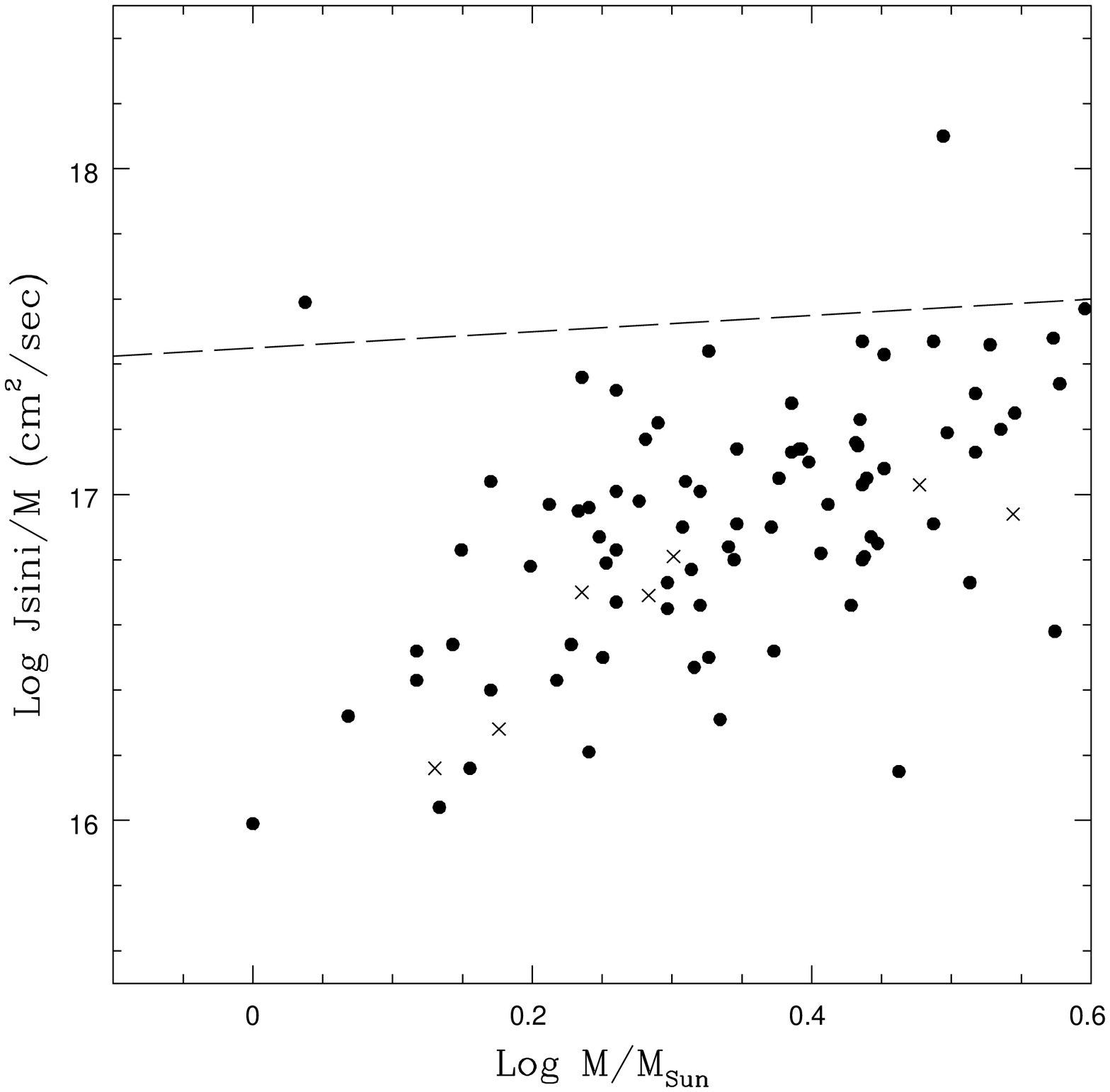} 
\figcaption{
(a) 
Values of specific angular momentum ({\it Jsini/M}) as a 
function of mass for stars on PMS convective tracks. Filled circles 
represent Orion stars in the current sample that are on convective 
tracks; open squares are similarly analyzed data from the study by
Rhode et al (2001) of primarily lower mass stars in the ONC;
filled triangles represent Orion stars with T$_{eff}$ \texttt{>} 10,000 K, 
which are already on the main sequence. The solid line shows 
the specific angular momentum corresponding to rotation at the 
breakup velocity along the PS birthline. The break in this curve 
occurs in the mass range where the birthline changes from intersecting 
convective tracks (M  \texttt{<} 3 M$_\odot$) to intersecting radiative 
tracks (M  \texttt{>} 3.5 M$_\odot$).  The dashed line is a fit by 
eye to the upper bound of the data and has slope 0.25.
(b)
The specific angular momentum of stars that have completed 
the convective phase of evolution. Filled circles represent stars 
in Orion that are either on radiative PMS tracks or on the ZAMS. 
Stars with T$_{eff}$ \texttt{>} 10000 K and the dashed line
are repeated from Figure 3a. Crosses represent the average
values of {\it Jsini/M} for field stars taken from data in the literature 
(see text for references).  
(c)  
An enlargement of the region of Figure 3b for masses
in the range 1-3 M$_\odot$.  Note the downturn in {\it Jsini/M}
for both the Orion and
field stars for masses less than about 2 M$_\odot$.  This downturn is not seen 
in convective PMS stars in Figure 3a.
}
\end{figure}

\clearpage
\begin{figure}
\plotone{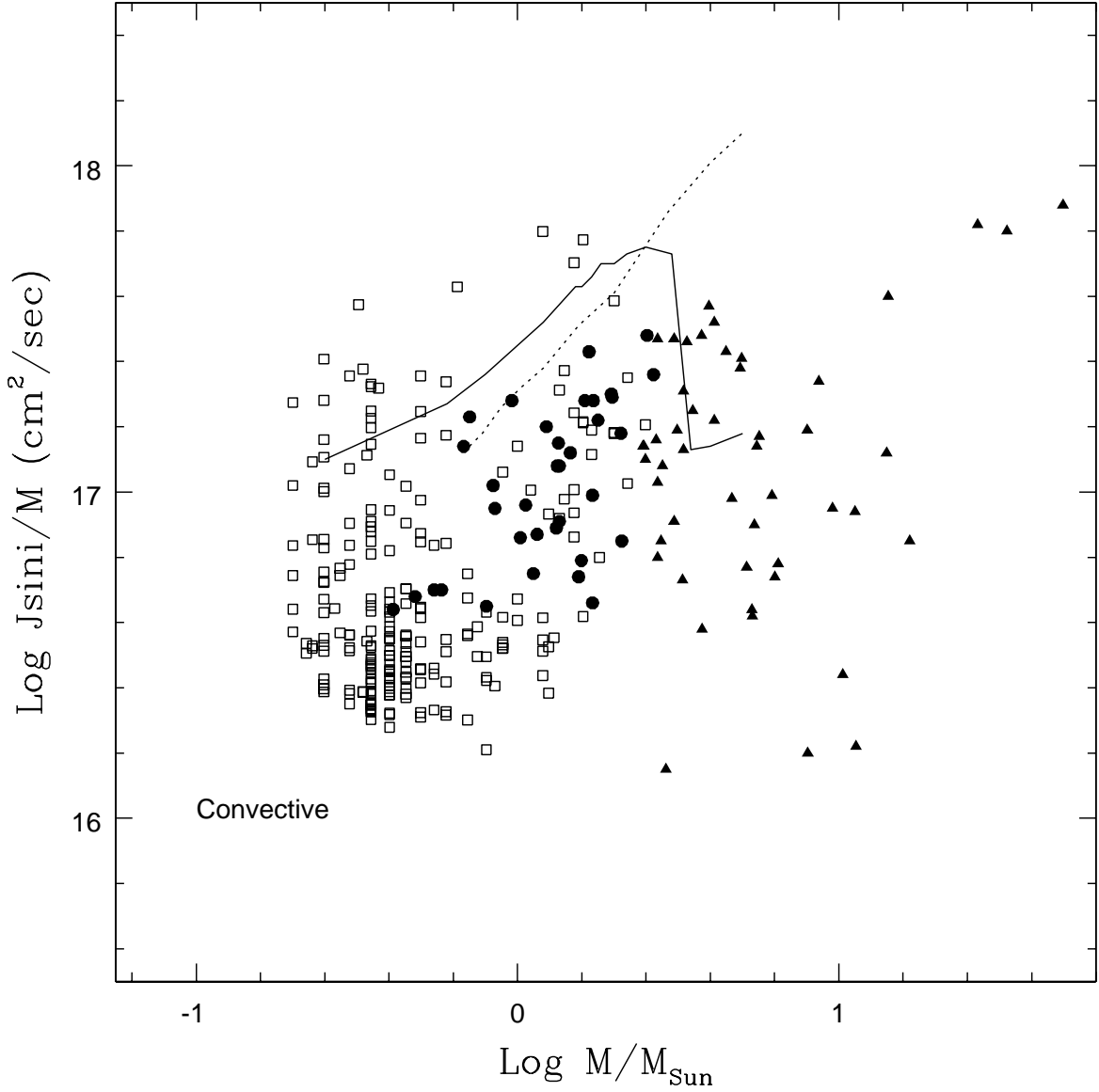}
\figcaption{
Predicted values of J/M along the 
PS birthline (solid line) and the BM birthline (dotted line) assuming
that the rotation of protostars is locked to their disks at least until 
they are released on the birthline. The data are the same as shown in
Figure 3a.  The break at 3 M$_\odot$ in the curve for the PS birthline
occurs where the birthline changes from intersecting convective tracks to
intersecting radiative tracks.  This transition takes place at 
M  \texttt{>} 5 M$_\odot$ for the higher accretion rates used
to calculate the BM birthline.
}
\end{figure}

\clearpage
\begin{figure}
\plotone{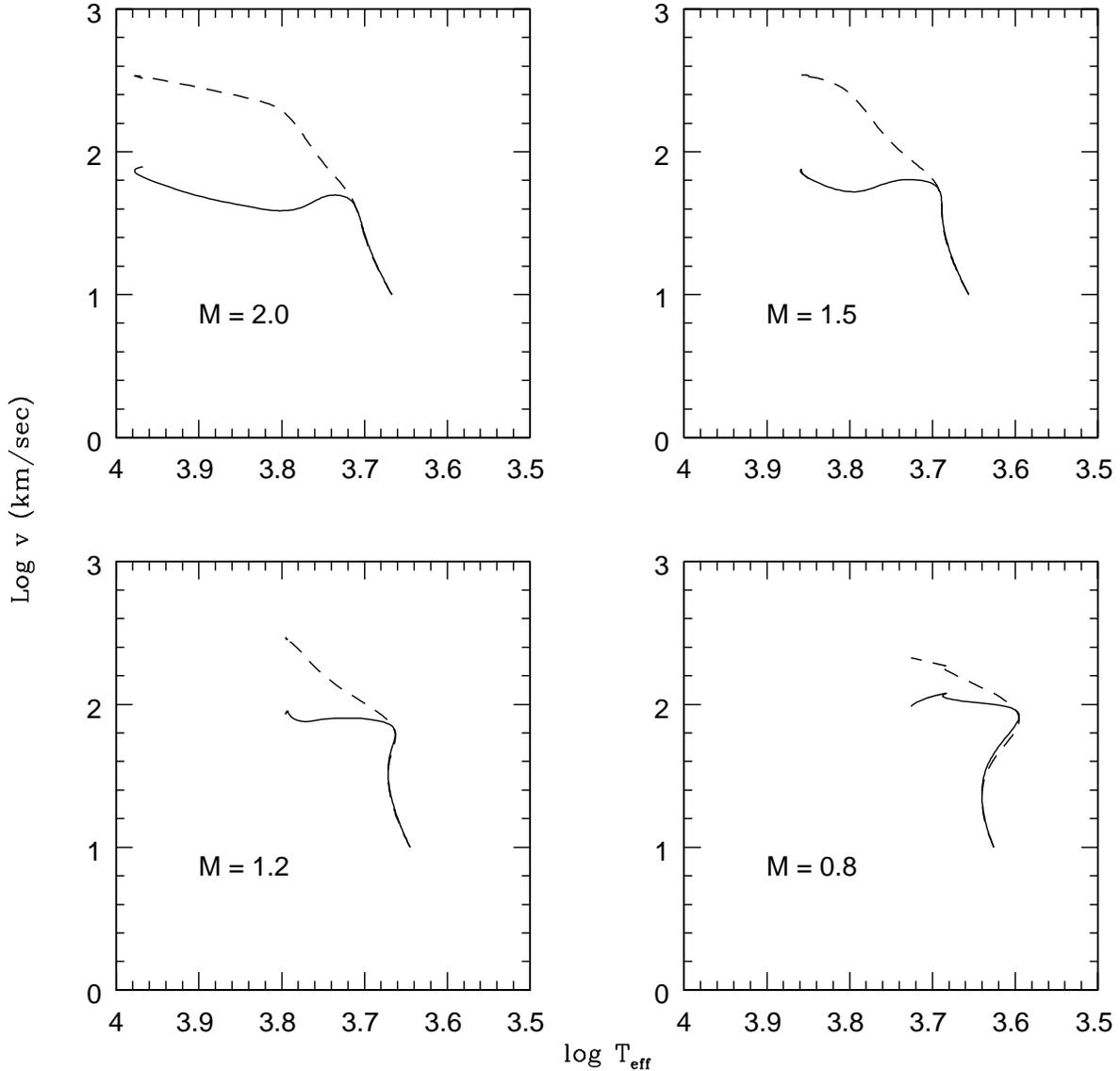}
\figcaption{
Changes from an assumed starting rotation of 10 km/s predicted by the SFRI 
models on the assumptions that: 1) there is no radial exchange of angular
momentum and angular momentum is conserved in shells (solid lines); and 2)
that stars rotate as solid bodies (dashed lines).  
The results are shown for four different masses. 
The stars initially evolve at nearly constant T$_{eff}$ toward higher 
rotational velocities as they evolve down their convective tracks. 
The transition from convective to radiative tracks coincides 
with the transition from evolution at nearly constant temperature 
to evolution toward higher temperatures but with relatively small 
changes in rotation. These calculations terminate when the stars 
reach the ZAMS. Note that the predictions of the two models are 
essentially identical for stars evolving along convective tracks. 
For the two higher mass models, the tracks are nearly parallel 
once the stars become fully radiative at about log T$_{eff}$ = 3.8, 
meaning that the fractional spinup is the same for the two cases 
along radiative tracks. However, the spinup predicted for solid 
body rotation during the transition from the convective to the 
radiative tracks is much larger than observed.
}
\end{figure}

\clearpage
\begin{figure}
\plotone{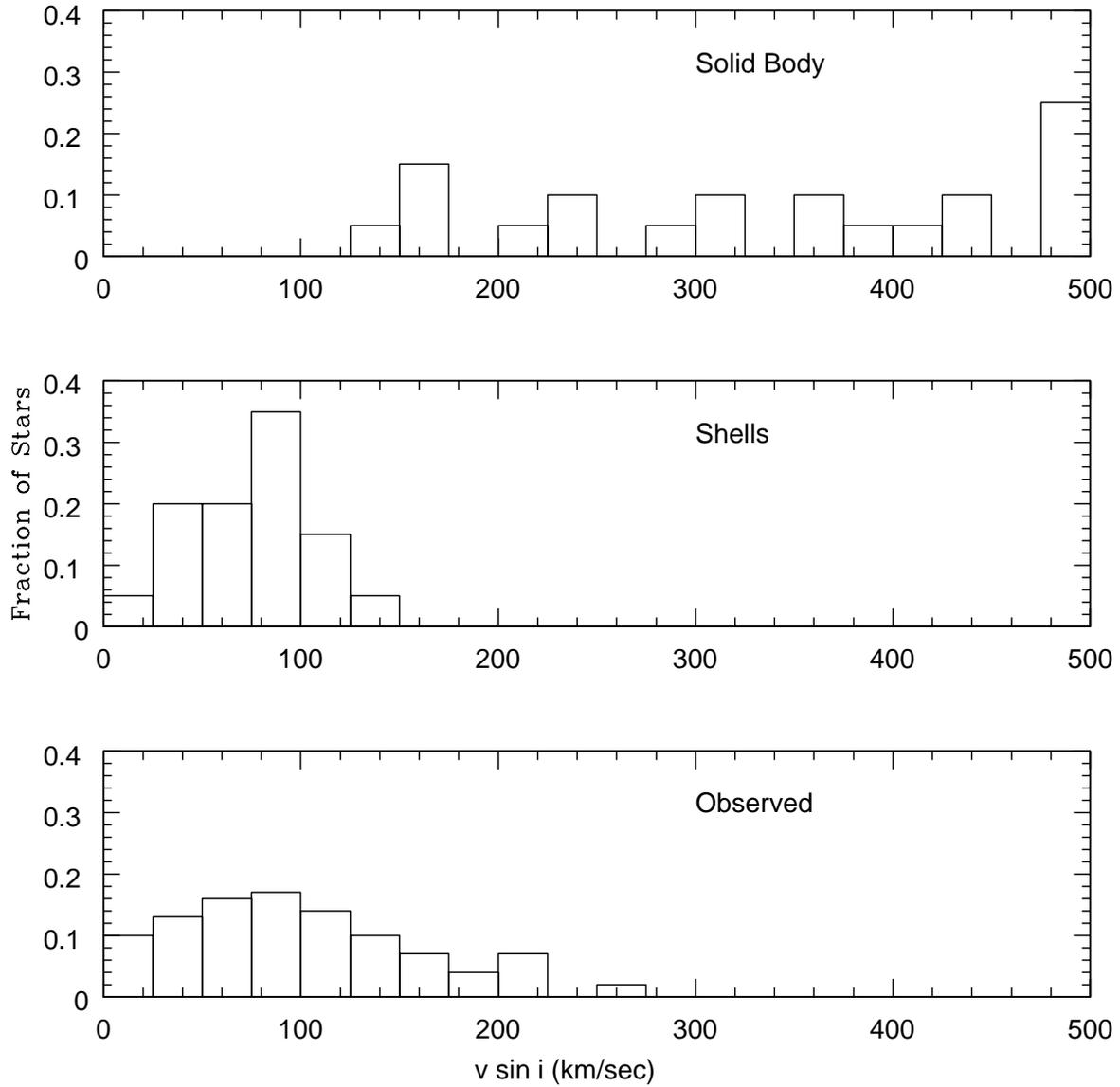}
\figcaption{
Observed and predicted ZAMS rotational velocities. 
The top panel shows the velocities expected on the ZAMS if the 
20 stars on convective tracks in the current sample with M \texttt{>} 
1.3 M$_{\odot}$ conserve angular momentum as solid bodies during their 
subsequent evolution to the main sequence; all of the stars predicted 
to have rotations in excess of 500 km/s are plotted in the 
rightmost box. The middle panel shows the predicted velocities 
if angular momentum is conserved in shells. The bottom panel 
shows the actual velocities for 69 field stars close to the ZAMS 
with masses in the range 1.5-2 M$_{\odot}$ taken from the study of 
Wolff \& Simon (1997).
}
\end{figure}

\clearpage
\begin{figure}
\plotone{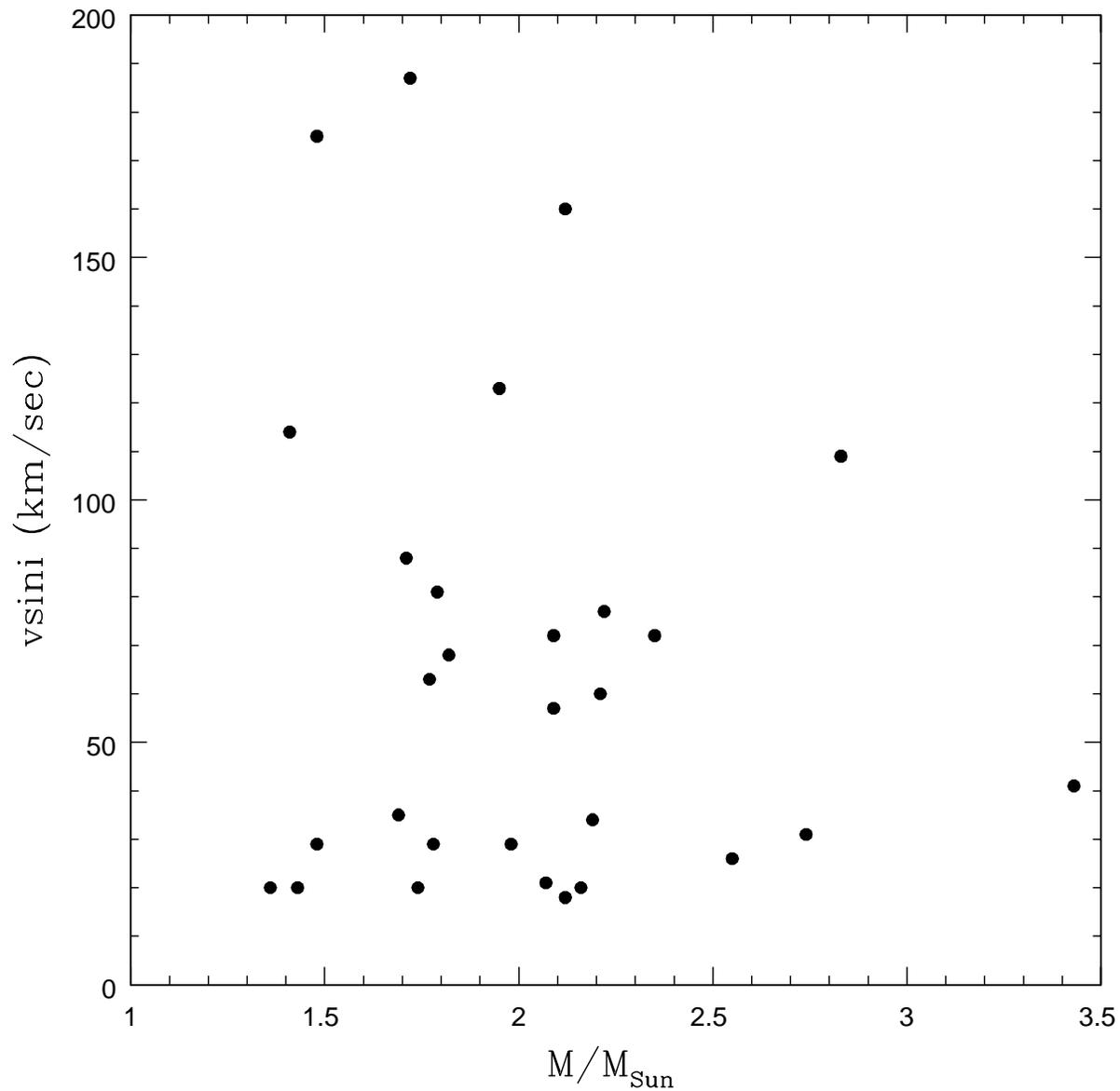}
\figcaption{
Apparent rotational velocities as 
a function of mass for stars in the temperature range 3.75 \texttt{<} log 
T$_{eff}$ \texttt{<} 3.85. All of 
these stars are on PMS radiative tracks (see Figures 2 and 5).
}
\end{figure}

\clearpage
\begin{figure}
\epsscale{0.8}
\plotone{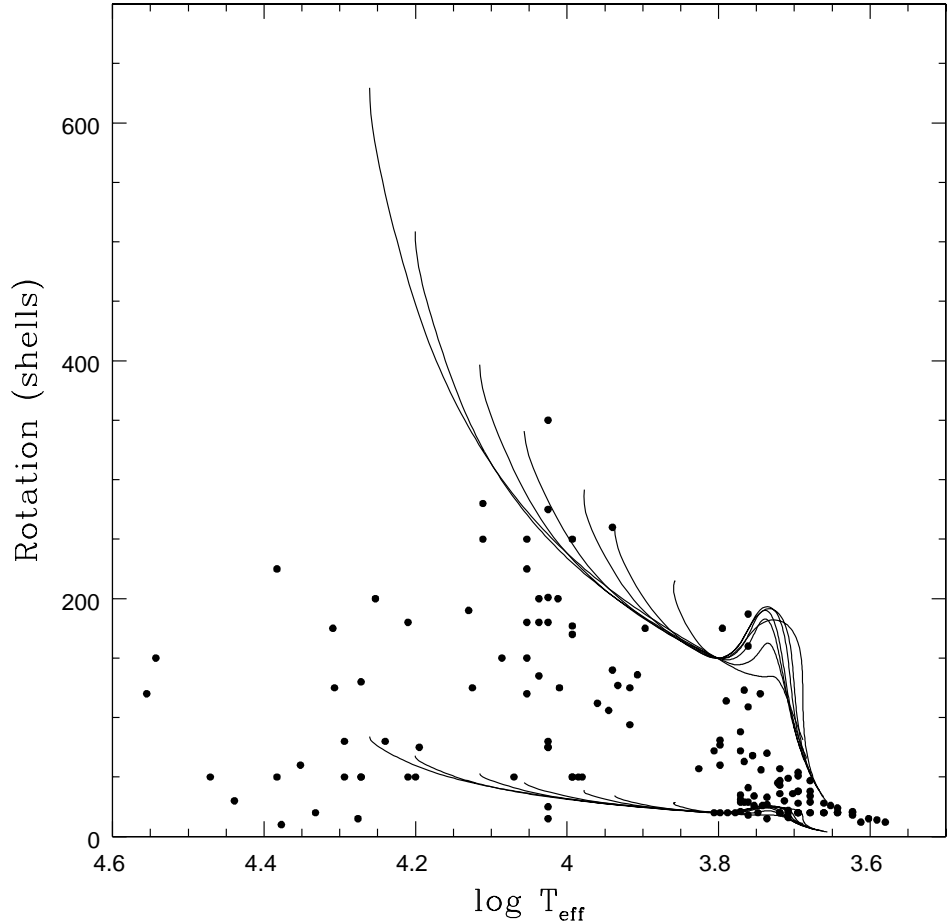}
\epsscale{1.0}
\figcaption{
Change in rotation predicted by 
the SFRI models on the assumption that angular momentum is conserved 
in shells 
The lower set of curves is for a rotation rate 
of 20 km/s at log T$_{eff}$ = 3.8, where the stars become fully 
radiative; the upper set is for a rotation rate of 150 km/s 
at log T$_{eff}$ = 3.8. The solid lines in each set of curves represent 
masses of 1.5, 1.8, 2, 2.5, 3, 4, and 5 M$_{\odot}$. These curves 
terminate when the stars reach the ZAMS. Filled circles represent 
data for the Orion stars in Table 2. Up to log 
T$_{eff}$ = 4.1, nearly all of the stars fall between the two sets 
of curves and rotation increases systematically with mass. 
This trend results from the fact that more massive stars traverse 
longer radiative tracks, contract more, and hence spin up more 
by the time they reach the main sequence. The trend breaks down, 
however, for stars with log T$_{eff}$ \texttt{>} 4.1 or masses larger 
than about 3 M$_{\odot}$. The maximum observed velocities for these 
more massive stars lie well below the values predicted if we 
assume that the stars traverse the full radiative track from 
T$_{eff}$ = 3.8 to the ZAMS. The probable explanation is that the 
birthline for masses greater than 3.5 M$_{\odot}$ for the models chosen 
here crosses the radiative portion of the evolutionary tracks. 
The distance traversed along radiative tracks from the birthline 
to the ZAMS, and hence the amount of spin up, decreases with 
increasing mass for masses greater than 3.5 M$_{\odot}$. 
We have also projected the predictions of \textit{vsini} backwards to 
the convective phase, again on the assumption that angular momentum 
is conserved in shells. The fact that the convective PMS stars 
fall within the bands so derived is another indication that the 
assumption that angular momentum is conserved in shells provides 
a good description of the data. 
}
\end{figure}

\clearpage
\begin{figure}
\plotone{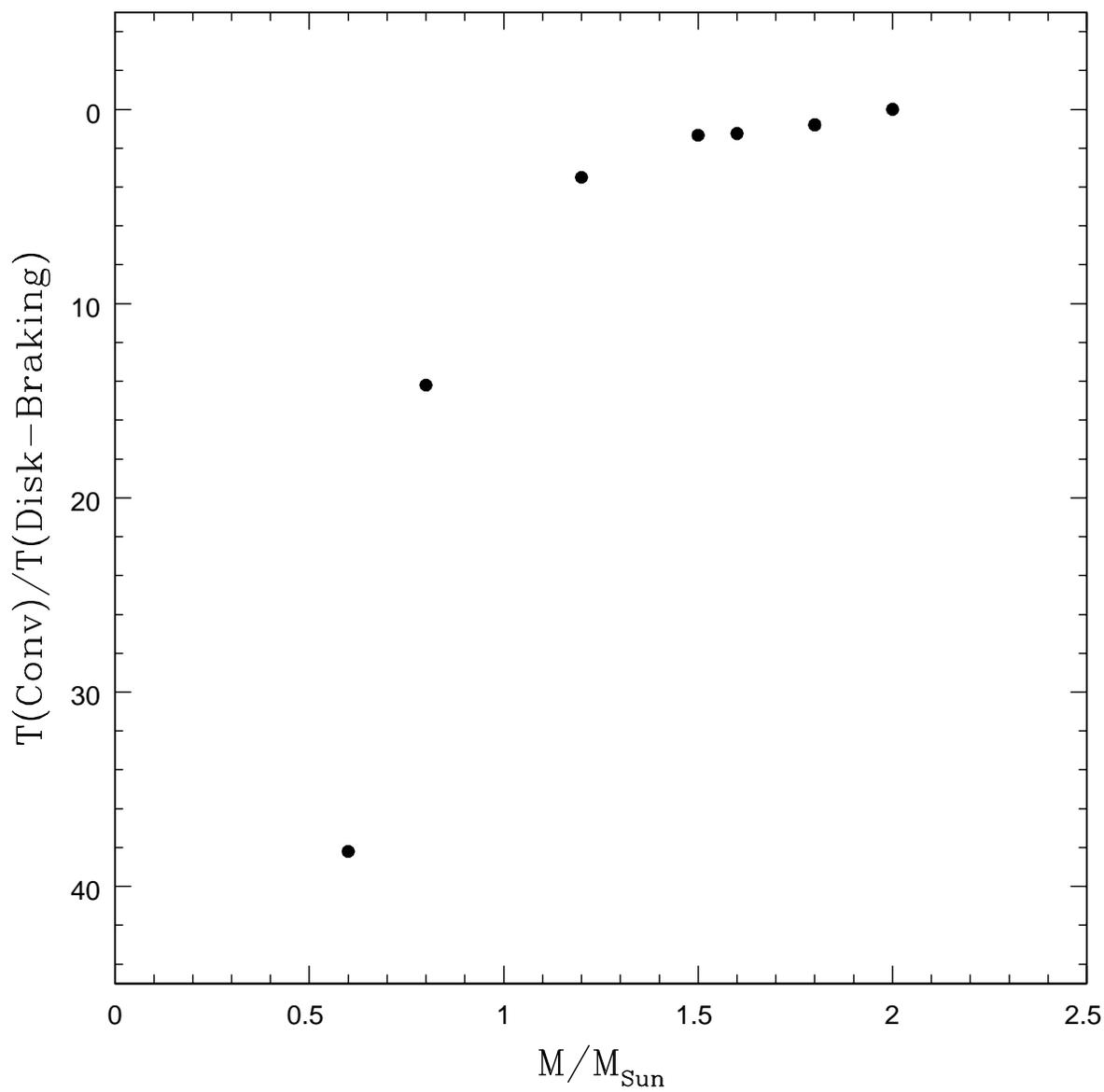}
\figcaption{
The ratio of the time spent on the convective track 
to the time scale for disk-braking estimated by the formulation of 
Hartmann (2002). Stars with masses close to 2 M$_{\odot}$ evolve too 
rapidly to shed much angular momentum during this phase of evolution.
}
\end{figure}

\end{document}